\documentclass[aps,prb,twocolumn,showpacs,superscriptaddress]{revtex4}
\usepackage{amssymb,amsmath}
\usepackage{graphicx}
\usepackage{multirow}
\usepackage{appendix}
\usepackage{mathtools}
\usepackage{amsmath}
\usepackage{xcolor}
\usepackage{siunitx}
\usepackage{comment}
\usepackage{hyperref}
\usepackage{footnote}
\usepackage{array}
\usepackage{multirow}
\usepackage{dsfont}
\usepackage{tabularray}

\graphicspath{{./img/}}

\begin{document}

\title{Variational first-principles approach to self-trapped polarons}

\author{Vasilii Vasilchenko}
\affiliation{European Theoretical Spectroscopy Facility, Institute of Condensed Matter and Nanosciences, Universit\'{e} catholique de Louvain, Rue de l'observatoire 8, bte L07.03.01, B-1348 Louvain-la-Neuve, Belgium}
\author{Matteo Giantomassi}
\affiliation{European Theoretical Spectroscopy Facility, Institute of Condensed Matter and Nanosciences, Universit\'{e} catholique de Louvain, Rue de l'observatoire 8, bte L07.03.01, B-1348 Louvain-la-Neuve, Belgium}
\author{Samuel Ponc\'e}
\affiliation{European Theoretical Spectroscopy Facility, Institute of Condensed Matter and Nanosciences, Universit\'{e} catholique de Louvain, Rue de l'observatoire 8, bte L07.03.01, B-1348 Louvain-la-Neuve, Belgium}
\affiliation{WEL Research Institute, avenue Pasteur 6, 1300 Wavre, Belgium}
\author{Xavier Gonze}
\affiliation{European Theoretical Spectroscopy Facility, Institute of Condensed Matter and Nanosciences, Universit\'{e} catholique de Louvain, Rue de l'observatoire 8, bte L07.03.01, B-1348 Louvain-la-Neuve, Belgium}

\date{\today}

\begin{abstract}
The behavior of charge carriers in polar materials is governed by electron-phonon interactions, which affect their mobilities via phonon scattering and may localize carriers into self-induced deformation fields, forming self-trapped polarons.
We present a first-principles study of self-trapped polaron formation in paradigmatic polar semiconductors and insulators using the variational polaron equations framework and self-consistent gradient optimization.
Our method incorporates long-range corrections to the electron-phonon interaction, essential for finite-size systems.
We demonstrate how the variational approach enables the identification of multiple polaronic states and supports the analysis of polarons with arbitrarily large spatial extent via energy filtering.
The potential energy surfaces of the resulting polarons exhibit multiple local minima, reflecting distinct, symmetry-broken polaronic configurations in systems with degenerate band edges.
Our findings align with previous theoretical studies and establish a robust foundation for future ab initio studies of polarons, especially those employing variational methods.
\end{abstract}

\pacs{71.38.-k, 78.20.Bh}

\maketitle

\section{Introduction}
\label{sec:intro}

Materials with moderate to strong electron-phonon coupling (EPC) can give rise to the formation of polarons.
A polaron is a quasiparticle formed when a particle couples with bosons, such as lattice vibrations, becoming associated with a dressing cloud.
Although the concept applies broadly to different particle types, the notion of polaron originated from studies on the autolocalization of an additional electron or hole in ionic crystals~\cite{landau_electron_1933, pekar_local_1946, Landau1948}, and phonon-dressed electron and hole polarons remain the primary focus of most research.
In fact, the formation of such polarons has been observed in a wide range of materials~\cite{alexandrov_polarons_2007, franchini_polarons_2021},
both computationally~\cite{guster_large_2023, LafuenteBartolome2024, Hassani2025}
and experimentally~\cite{rene_de_cotret_direct_2022, shi2024formationanisotropicpolaronsantimony},
with new methodologies emerging to investigate and characterize these quasiparticles~\cite{sio_ab_2019, sio_polarons_2019, lee_facile_2021, vasilchenko_variational_2022, lafuente-bartolome_ab_2022, lafuente-bartolome_unified_2022,  falletta_polarons_2022, falletta_many-body_2022, sio_polarons_2023}.
The present work focuses on polarons of this kind.

Depending on the strength of the EPC, electron- and hole-polaron formation in semiconductors and insulators falls into two limiting regimes: weak coupling and strong coupling.
In the weak-coupling regime, the delocalized (Bloch-type) charged particle induces fluctuations in the phonon cloud, leading to polaronic effects such as electronic band-gap renormalization and effective mass enhancement.
In the strong-coupling regime, the charge carrier undergoes self-trapping in the induced deformation field, forming a localized polaronic state within the band gap.
This localization significantly alters charge transport properties, transitioning from band-like transport to hopping transport~\cite{Geneste2017, chang_intermediate_2022, Faletta2023, Birschitzky2024}.
Additionally, an intermediate regime exists where features of both weak and strong coupling emerge, requiring treatment beyond conventional approximations.

In this work, we focus on the variational treatment of self-trapped electrons and holes in polar materials. 
From a theoretical perspective, self-trapping has been addressed using various approaches, including model solutions~\cite{frohlich_xx_1950, frohlich_electrons_1954, holstein_studies_1959, holstein_studies_1959-1, guster_frohlich_2021, vasilchenko_polarons_cubic_2024},
density functional theory (DFT) with charged supercell relaxation~\cite{Sadigh2015, vasilchenko_small_2021, falletta_polarons_2022, falletta_many-body_2022},
and more recently, several \textit{ab initio} primitive cell methods~\cite{sio_ab_2019, sio_polarons_2019, lee_facile_2021, vasilchenko_variational_2022, lafuente-bartolome_ab_2022, lafuente-bartolome_unified_2022}.
These methods describe the autolocalization of a unit charge within an adjustable deformation field in a medium large enough to accommodate a polaron.
This adjustment typically involves a self-consistent field (SCF) cycle, where a localized charge induces a corresponding lattice deformation, which in turn determines the next lower-energy configuration.
Depending on the spatial extent of the polaron relative to the atomic length scale of a system, polarons are classified as small~\cite{falletta_many-body_2022, falletta_polarons_2022}, medium~\cite{Hassani2025}, or large~\cite{rene_de_cotret_direct_2022, guster_large_2023}.
In the latter cases, the size of the system required to accommodate the polaron makes approaches based on supercell relaxation computationally prohibitive.
Such cases can be effectively addressed using models or primitive-cell methodologies.

However, all SCF-based approaches face a fundamental challenge: while the iterative approach aims to converge to a stable polaronic configuration, it does not guarantee that the solution corresponds to the global minimum of the polaron energy surface.
A similar issue is known in spin-dependent Kohn–Sham DFT, where different metastable spin states can prevent finding the true ground state~\cite{davis_global_opt_2014, woods_computing_scf_2019}.
Likewise, in polaronic SCF methods, the initialization of the SCF cycle can lead to different local minima, resulting in higher energy bound polaron states, and degenerate solutions with equivalent energy can be overlooked.
Additionally, spontaneous symmetry-breaking effects often occur during polaron formation~\cite{vasilchenko_polarons_cubic_2024},
where the ground-state polaron may not retain the original symmetry of the system, favoring lower-symmetry configurations.
These challenges arise even in simplified polaron models~\cite{guster_frohlich_2021, vasilchenko_polarons_cubic_2024} and become more complex in first-principles calculations~\cite{Frodason2020, Faletta2023}.

In this work, we study these phenomena using the variational approach to \textit{ab initio} self-trapped polaron formation, and present it as a powerful method to overcome such challenges in real materials.
Specifically, the variational polaron equations (VarPEq) framework~\cite{vasilchenko_variational_2022}, implemented within the ABINIT software~\cite{Gonze2016, gonze_abinitproject_2020} is used to investigate the formation of electron and hole polarons in various materials and obtain non-trivial degenerate localized polaronic states.
The nature of this degeneracy is analyzed and connected to the symmetry-breaking of polaron formation.
Additionally, we explore how to accelerate convergence towards polaronic solutions by mitigating finite-size effects inherent to the problem.
Finally, the VarPEq capability to efficiently compute polarons with arbitrarily large localization lengths is demonstrated.

Sec.~\ref{sec:theory} provides the theoretical background and introduces the problem of self-trapped polaron formation. 
We then describe a technique for computing nontrivial degenerate polaronic solutions and a method to account for long-range effects, which improves convergence.
We further establish the classification of polarons by their spatial extent, and discuss how arbitrarily large polarons can be addressed within our method.
Sec.~\ref{sec:setup} discusses the computational details of polaron calculations in LiF, MgO, Li$_2$O$_2$, wurtzite, and zincblende GaN.
In Sec.~\ref{sec:results}, the obtained self-trapped polaron solutions are presented and compared with previously reported theoretical results, highlighting the applicability and limitations of the methodology.

\section{Methodology}
\label{sec:theory}

\subsection{Self-trapped polarons}

The addition of an electron or a hole distorts the surrounding lattice, which, in turn, generates a polarization field that acts as a potential well, leading to the autolocalization of the charge.  
In that framework, the polaron binding energy, $E_\mathrm{pol}$, is defined as the energy difference between the charged system in the distorted (polaronic) geometry and the undistorted one,
\begin{equation}
    E_\mathrm{pol} =
    E \left( N \pm 1, \Delta \boldsymbol{\tau} \right)
    -
    E \left( N \pm 1, \Delta \boldsymbol{\tau} \equiv 0 \right),
\end{equation}
where $E$ denotes the total energy of a system with $N \pm 1$ electrons, and $\Delta \boldsymbol{\tau} = \left\{ \Delta \tau_{\kappa \alpha p} \right\}$ represents the lattice distortion as atomic displacements from the equilibrium positions. 
Here, $\kappa$ indexes the atoms in the unit cell $p$ and $\alpha$ denotes the displacement directions. 
A stable polaronic configuration corresponds to $E_\mathrm{pol} < 0$.

\begin{figure}[t]
    \centering
    \includegraphics[width=1.\linewidth]{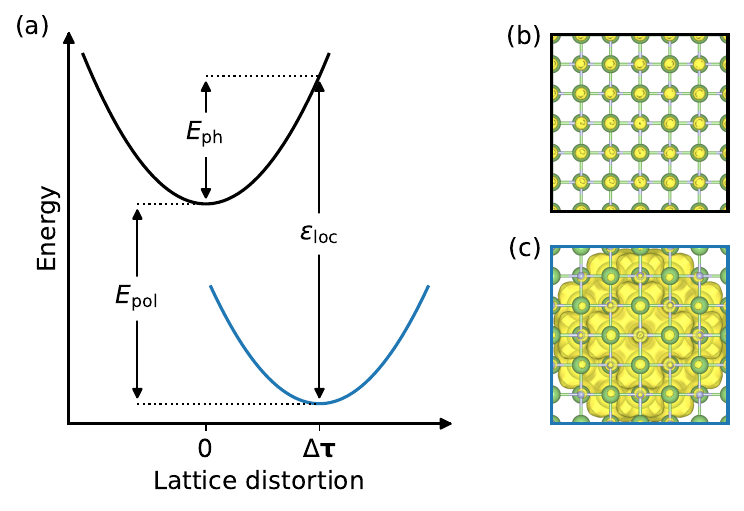}
    \caption{
    Configurational diagram showing the energy of a system with $N \pm 1$ electrons as a function of lattice distortion.  
    Black and blue curves in panel (a) represent the undistorted and polaronic configurations, respectively.  
    $E_\mathrm{pol}$ denotes the polaron formation energy, while $E_\mathrm{ph}$ and
    $\varepsilon_{\rm loc}$ correspond to the lattice and charge localization contributions, respectively.  
    Panels (b) and (c) illustrate charge density distributions for an extra electron in LiF, corresponding to the minima of the undistorted and polaronic configurations.
    }
    \label{fig:conf}
\end{figure}

The configurational diagram in Fig.~\ref{fig:conf} schematically illustrates this process.
In this representation, $E_\mathrm{pol}$ can be decomposed into two contributions: the phonon energy $E_\mathrm{ph}$, arising purely from lattice distortion, and the charge localization energy $\varepsilon_{\rm loc}$, which accounts for both electronic contributions and EPC.  
Physically, the absolute value of the latter term corresponds to the energy of the localized polaronic state inside the band gap, measured from the valence band maximum (VBM) for a hole polaron and from the conduction band minimum (CBM) for an electron polaron.

A bound polaron exhibits localized lattice deformation $\Delta \boldsymbol{\tau}$ and charge density $\rho(\mathbf{r})$.
This localization is characterized by a length scale $a_{\rm pol}$, which determines whether the polaron is classified as small, medium, or large, relative to a characteristic length scale of the system.
The precise definition of $a_{\rm pol}$ and the classification criteria adopted in this work are established in Sec.~\ref{sec:aloc}.

Regardless of the spatial extent, the interaction between the charge and atomic displacements can be described in the adiabatic approximation, where lattice fluctuations are neglected, and the charge distribution is assumed to adjust instantaneously to the induced polarization.  
While non-adiabatic effects on the band renormalization due to electron-phonon interactions can also be incorporated, their simultaneous treatment with adiabatic effects requires advanced techniques~\cite{lafuente-bartolome_ab_2022, lafuente-bartolome_unified_2022}.
In this work, we focus exclusively on the self-trapped electron and hole polaron formation with arbitrarily large localization lengths within the adiabatic regime.

\subsection{Variational polaron equations}

In the adiabatic regime, a variational expression for $E_\mathrm{pol}$ is available.
Variational approaches have consistently shown their effectiveness in solving various polaron models~\cite{Landau1948, j_miyake_strong-coupling_1975, j_miyake_ground_1976, vasilchenko_variational_2022, vasilchenko_polarons_cubic_2024}.
In this study, we are relying on the VarPEq methodology.
This section summarizes the main features of the method, while a detailed discussion of its derivation is presented elsewhere~\cite{vasilchenko_variational_2022}.
Throughout the work, we use the Hartree atomic units: $\hbar = m_e = |e| = 1$.

The foundational principles of the VarPEq methodology originate from the work of Sio \textit{et al.}~\cite{sio_ab_2019, sio_polarons_2019} who reformulated the problem of \textit{ab initio} polaron formation from the supercell representation to the reciprocal space and derived an effective polaron Hamiltonian, parametrized from first principles.
Using the same premises but without relying on an effective Hamiltonian, the VarPEq method approaches the problem with a variational expression for the polaron binding energy~\cite{vasilchenko_variational_2022}.
The charged particle is assumed to localize within a Born-von Karman supercell, consisting of $N_p$ unit cells and represented by the corresponding $\mathbf{k}$-point sampling of the Brillouin zone (BZ).
The polaron wavefunction $\phi(\mathbf{r})$ in real space is constructed from the undistorted Bloch wavefunctions $\psi_{n\mathbf{k}}(\mathbf{r})$ as
\begin{equation}\label{eq:varphi}
    \phi (\mathbf{r})
    =
    \frac{1}{\sqrt{N_p}}
    \sum_{n\mathbf{k}} A_{n\mathbf{k}} \psi_{n\mathbf{k}} (\mathbf{r}),
\end{equation}
where summation runs over the states $n\mathbf{k}$ participating in the polaron formation.
The coefficients $A_{n\mathbf{k}}$ will be determined later.
The electronic charge density directly comes from the modulus square of the polaron wavefunction:
\begin{equation}\label{eq:rho}
    \rho(\mathbf{r}) = | \phi (\mathbf{r})|^2.
\end{equation}

Similarly, the induced lattice distortion is expressed in terms of atomic displacements:
\begin{equation}\label{deltatau}
    \Delta \tau_{\kappa \alpha p} = - \dfrac{2}{N_p} \sum_{\mathbf{q}\nu}
    B^*_{\mathbf{q}\nu} \Big[ \dfrac{1}{2M_{\kappa}\omega_{\mathbf{q}\nu}}\Big]^{1/2} \!\!  e_{\kappa \alpha \nu}(\mathbf{q}) e^{i\mathbf{q}\cdot\mathbf{R}_p}.
\end{equation}
Here, $M_{\kappa}$ denotes the atomic mass, $e_{\kappa\alpha\nu} (\mathbf{q})$ are the phonon eigenmodes with eigenvalues $\omega_{\mathbf{q}\nu}$, and $\mathbf{R}_p$ is the position vector of a unit cell $p$.

The variational subspace consists of the electronic and vibrational coefficients $\boldsymbol{A} \equiv \left\{ A_{n\mathbf{k}} \right\}$ and $\boldsymbol{B} \equiv \left\{ B_{\mathbf{q}\nu} \right\}$, from which the charge distribution and the induced deformation field are obtained, respectively.
Due to the underlying adiabatic approximation, these coefficients are treated independently.

The polaron formation is then characterized by parameters defined in the BZ of a single unit cell: the electronic band structure $\varepsilon_{n\mathbf{k}}$, phonon dispersion $\omega_{\mathbf{q}\nu}$, and electron-phonon matrix elements $g_{mn\nu}(\mathbf{k, q})$.
For electron and hole polarons, electronic bands are assumed to belong to the conduction and valence manifolds, respectively.
The variational expression for the binding energy is given by~\cite{vasilchenko_variational_2022}
\begin{equation}\label{eq:varpeq}
    E_\mathrm{pol} \left( \boldsymbol{A}, \boldsymbol{B}  \right) = E_\mathrm{el} \left( \boldsymbol{A} \right)
    + E_\mathrm{ph} \left( \boldsymbol{B}  \right)
    + E_\mathrm{e\text{-}ph} \left( \boldsymbol{A}, \boldsymbol{B}  \right),
\end{equation}
where 
\begin{align}
    E_\mathrm{el} \left( \boldsymbol{A} \right)
    =& \frac{1}{N_p} \sum_{n\mathbf{k}}
    | A_{n\mathbf{k}} |^2
    \left( \varepsilon_{n\mathbf{k}} -
    \varepsilon_{\rm loc} \right) + \varepsilon_{\rm loc} \label{eq:varpeq:el} \\
    E_\mathrm{ph} \left( \boldsymbol{B} \right) =& \frac{1}{N_p} \sum_{\mathbf{q}\nu}
    | B_{\mathbf{q}\nu} |^2 \omega_{\mathbf{q}\nu}, \label{eq:varpeq:ph} \\ 
    E_\mathrm{e\text{-}ph} \left( \boldsymbol{A}, \boldsymbol{B} \right) =& - \frac{1}{N_p^2} \sum_{mn\nu \mathbf{kq}} \nonumber \\
    & \times \! A^*_{m\mathbf{\mathbf{k+q}}} B^*_{\mathbf{q}\nu} g_{mn\nu}(\mathbf{k, q})
    A_{n\mathbf{\mathbf{k}}} \!+\! \mathrm{(c.c)},\label{eq:varpeq:elph}
\end{align}
represent electronic, phonon, and electron-phonon contributions to the binding energy, respectively.
The phonon-related terms in $\mathbf{q}$-space are sampled to incorporate all possible transitions from a state $\psi_{n\mathbf{k}}$ to a state $\psi_{m\mathbf{k}'}$, which obey the momentum conservation relation:
\begin{equation}
    \mathbf{q} = \mathbf{k}' - \mathbf{k} + \mathbf{G},
\end{equation}
where $\mathbf{G}$ is an Umklapp vector mapping $\mathbf{q}$ back into the first BZ.
For simplicity, we adopt identical $\Gamma$-centered uniform $\mathbf{k}$-point and $\mathbf{q}$-point meshes.

The polaron binding energy $E_\mathrm{pol} \left( \boldsymbol{A}, \boldsymbol{B} \right)$ is optimized subject to the normalization constraint of a single charge in the system:
\begin{equation}\label{eq:anorm}
    \left<
    \phi | \phi
    \right>
    =
    \frac{1}{N_p}
    \sum_{n\mathbf{k}} | A_{n\mathbf{k}} |^2 = 1.
\end{equation}
This constraint is included through the Lagrange multiplier $\varepsilon_{\rm loc}$ in Eq.~\eqref{eq:varpeq:el}, which corresponds to the charge localization contribution shown in Fig.~\ref{fig:conf}.
The energy of the forming localized polaronic state is measured relative to the band edge.

So far, we have not specified the origin of the parameters 
$\varepsilon_{n\mathbf{k}}$, $\omega_{\mathbf{q}\nu}$, and $g_{mn\nu}(\mathbf{k, q})$, defining the electronic, phonon, and electron-phonon contributions to the polaron formation.
In principle, the VarPEq method is agnostic to the origin of these parameters.
For example, it has been applied to the generalized Fr\"ohlich model, where parameters were derived analytically~\cite{vasilchenko_variational_2022, guster_large_2023, vasilchenko_polarons_cubic_2024}.
In the present work, these come from first-principles DFT calculations.
Details on this parametrization for specific materials are provided in Sec.~\ref{sec:setup}.

\subsection{Self-consistent gradient optimization}
\label{sec:scf}

One of the advantages of the VarPEq method is that it provides an explicit variational expression for $E_\mathrm{pol}$.
This enables the use of efficient gradient-based optimization techniques for finding polaronic solutions.
In this section, we outline the set of core self-consistent equations required for gradient approaches within the framework.

The variational problem defined by Eqs.~\eqref{eq:varpeq}-\eqref{eq:varpeq:elph} adjusts the polaronic configuration to minimize $E_\mathrm{pol}$.
For a given charge distribution originating from a set $\boldsymbol{A}$, the condition on the phonon gradient is the following:
\begin{align}
    0 \equiv &  \nabla_{B_{\mathbf{q}\nu}} E_\mathrm{pol} \nonumber \\
    =& \frac{2}{N_p} B_{\mathbf{q}\nu} \omega_{\mathbf{q}\nu} -
    \frac{2}{N_p^2}
    \sum_{nm\mathbf{k}}
    A^*_{m\mathbf{k+q}}
    g_{mn\nu}(\mathbf{k,q})
    A_{n\mathbf{k}}.
\end{align}
This yields the optimal vibrational coefficients:
\begin{equation}\label{eq:phgrad}
    B_{\mathbf{q}\nu} \left( \boldsymbol{A} \right)
    = \frac{1}{N_p}\sum_{mn\mathbf{k}}
    A^*_{m\mathbf{k+q}}
    \frac{g_{mn\nu}(\mathbf{k, q})}{\omega_{\mathbf{q}\nu}}
    A_{n\mathbf{k}}.
\end{equation}
At this stage, having determined $\boldsymbol{B}$ for a given set of $\boldsymbol{A}$ coefficients, the gradient of $E_\mathrm{pol}$ with respect to the electronic degrees of freedom is obtained as
\begin{multline}\label{eq:elgrad}
    D_{n\mathbf{k}} \bigl( \boldsymbol{A}, \boldsymbol{B}, \varepsilon_{\rm loc} \bigr)
    \equiv \nabla_{A_{n\mathbf{k}}} E_{\rm pol} 
    = \frac{2}{N_p} A_{n\mathbf{k}}
    \left( \varepsilon_{n\mathbf{k}} - \varepsilon_{\rm loc} \right) \\
    - \frac{2}{N_p^2}
    \sum_{m\nu\mathbf{q}}
    \big( A_{m\mathbf{k-q}}
    B^*_{\mathbf{q}\nu} g_{nm\nu}(\mathbf{k-q}, \mathbf{q}) \\ 
    +
    A_{m\mathbf{k+q}}
    B_{\mathbf{q}\nu} g^*_{mn\nu}(\mathbf{k}, \mathbf{q})\big),
\end{multline}
where the charge localization energy $\varepsilon_{\rm loc}$ is computed from the $\nabla_{A_{n\mathbf{k}}} E_{\rm pol} \equiv 0$ condition on the gradient, and the normalization condition of Eq.~\eqref{eq:anorm}:
\begin{multline}\label{eq:eps}
    \varepsilon_{\rm loc}  = \frac{1}{N_p}
    \sum_{n\mathbf{k}} | A_{n\mathbf{k}} |^2 \varepsilon_{n\mathbf{k}} \\
    - \frac{1}{N_p^2} \sum_{\substack{mn\nu \\ \mathbf{kq}}}
    A^*_{m\mathbf{k+q}}
    B^*_{\mathbf{q}\nu} g_{mn\nu}(\mathbf{k}, \mathbf{q})
    A_{n\mathbf{k}} + \text{(c.c.)}.
\end{multline}

Equations~\eqref{eq:phgrad} and \eqref{eq:eps} describe the deformation field and the localization energy of the charge when the electronic coefficients are known.
Using the electronic gradient defined by Eq.~\eqref{eq:elgrad}, a step is taken in the direction of the steepest descent to update the electronic coefficients.
This step preserves the normalization condition of Eq~\eqref{eq:anorm} via line minimization
\begin{equation}\label{eq:linemin}
    \boldsymbol{A} \to  \boldsymbol{A}\cos{\theta} + \boldsymbol{D}_\perp \sin{\theta},
\end{equation}
where $\boldsymbol{D}_\perp$ is obtained by the orthonormalization of the gradient with its associated wavefunction
\begin{equation}
    \xi (\mathbf{r})
    =
    \frac{1}{\sqrt{N_p}}
    \sum_{n\mathbf{k}} D_{n\mathbf{k}} \psi_{n\mathbf{k}} (\mathbf{r}),
\end{equation}
such as 
\begin{equation}\label{eq:linemin-ort}
    \boldsymbol{D}_\perp =
    \boldsymbol{D}
    -
    \left<
    \phi
    |
    \xi 
    \right>
    \boldsymbol{A},
\end{equation}
and
\begin{equation}\label{eq:linemin-grad-norm}
    \left<
    \xi_\perp
    |
    \xi_\perp
    \right>
    =  
    \frac{1}{N_p}
    \sum_{n\mathbf{k}} | D_{\perp, n\mathbf{k}} |^2
    = 1.
\end{equation}
This process is iterated until convergence, which yields a solution.
The optimization for electron and hole polarons is symmetric, provided that for a hole polaron, the valence bands entering Eq.~\eqref{eq:varpeq:el} are flipped, ensuring that the problem remains minimization.

The self-consistency of this process reflects the autolocalization of the charge for localized (non-trivial) solutions, as the resulting deformation field ensures that the particle becomes self-trapped with an optimal charge distribution.
Indeed, a solution is found when both the phonon and electron gradients vanish.
We note that a delocalized (trivial) solution is obtained with $ A_{n\mathbf{k}} \equiv 0 $ for all states except at the band edges.
Equations~\eqref{eq:phgrad}-\eqref{eq:linemin} can be solved with various gradient-based optimization methods, and we choose the preconditioned conjugate gradient algorithm (PCG)~\cite{vasilchenko_variational_2022}.

In our implementation, the conjugate direction is calculated with the Polak-Ribi\`ere formula, which allows for flexible preconditioning~\cite{Notay2000}.
The preconditioner $\hat{P}$ comes naturally from the definition of the gradient in Eq.~\eqref{eq:elgrad} and is taken to be a diagonal positive-definite matrix, which depends on $\varepsilon_{\rm loc}$:
\begin{equation}\label{eq:pcond}
    P_{n\mathbf{k},n'\mathbf{k}'} \left( \varepsilon_{\rm loc} \right)
    =
    \delta_{n\mathbf{k},n'\mathbf{k}'}
    |\varepsilon_{n\mathbf{k}} + |\varepsilon_{\rm loc}||^{-1}.
\end{equation}
Note that it may also be modified to account for the long-range EPC, as detailed in Sec.~\ref{sec:lr}.

\subsection{Degenerate polaronic solutions}
\label{sec:multipol}

Beyond computational efficiency, explicit treatment of the gradient in the VarPEq approach facilitates the computation of additional polaronic solutions after an initial solution has been identified.  
Here, we address the problem of multiple polaronic states and outline how they can be systematically uncovered through gradient-based optimization within the variational framework.

The self-consistent optimization defined by Eqs.~\eqref{eq:phgrad}-\eqref{eq:eps} is sensitive to the initial choice of electronic coefficients, $\boldsymbol{A}$.
Already in the generalized Fr\"ohlich model with degenerate electronic bands, multiple polaronic solutions with distinct symmetry groups are found~\cite{vasilchenko_polarons_cubic_2024}.
Numerically, this means that different initial charge distributions may converge to different polaronic solutions.

Furthermore, considering the atomic structure of the supercell that hosts the polaron, any localized polaron with configuration $\boldsymbol{A}$ inherently has $N_p$ equivalent configurations of the form
\begin{equation}  
    \left\{ e^{i\mathbf{k \cdot R_p}} A_{n\mathbf{k}} \right\},  
\end{equation}  
which correspond to translations of the solution. 
These configurations are translationally equivalent, share the same formation energy, and collectively represent the same polaronic state.
\begin{figure}[t]
    \centering
    \includegraphics[width=1.\linewidth]{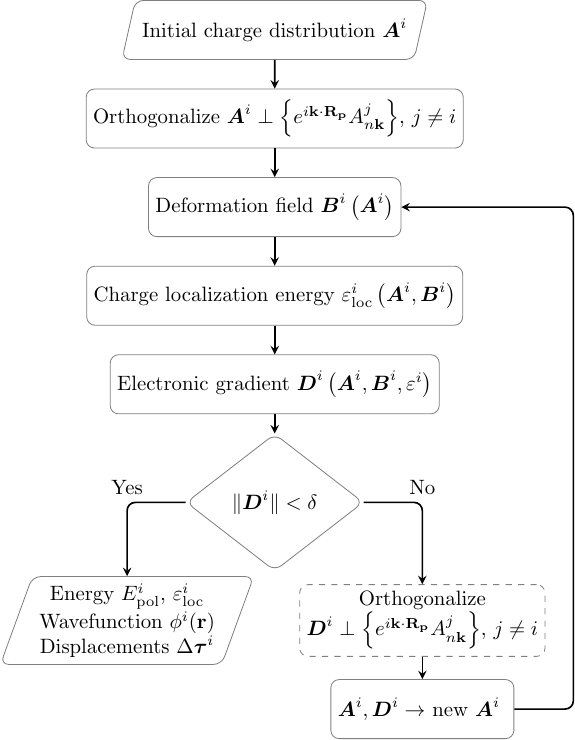}
    \caption{
    Flow diagram for the variational optimization of an $i$-th polaronic state.
    The orthogonality constraint to all prior solutions and their translated images is first enforced by orthogonalizing the initial charge distribution $\boldsymbol{A}^i$ against the previous ones, $\left\{ e^{i\mathbf{k \cdot R_p}} A^j_{n\mathbf{k}} \right\},~j \ne i$.
    This constraint is then preserved by applying a similar orthogonalization to the electronic gradient in the self-consistent cycle.
    The dashed frame of the gradient orthogonalization process indicates that the constraint can be removed if the process terminates at a constraint-induced stationary point.
    $\delta$ denotes the tolerance on the gradient norm.
    }
    \label{fig:ort_flow}
\end{figure}
These examples illustrate that the potential energy surface, $E_\mathrm{pol} \left( \boldsymbol{A}, \boldsymbol{B} \right)$, as defined in Eq.~\eqref{eq:varpeq}, can exhibit multiple minima corresponding to qualitatively distinct polaronic states, each possessing $N_p$ translationally equivalent images on the energy surface.

To identify inequivalent polaronic solutions, we use the following procedure.
When optimizing for the $i$-th polaronic state with configuration $\boldsymbol{A}^i$ and corresponding wavefunction $\phi^i$, we enforce it to be orthogonal to all previously obtained solutions $\boldsymbol{A}^j,~j < i$:
\begin{equation}\label{eq:ort-constr}
    \left<
    \phi^i | \phi^j
    \right>
    =
    \frac{1}{N_p}
    \sum_{n\mathbf{k}} {A^i}^*_{n\mathbf{k}} A^j_{n\mathbf{k}} = 
    \delta_{ij}.
\end{equation}
This is achieved by first orthogonalizing the electronic gradient to all prior solutions before the line minimization step of Eq~\eqref{eq:linemin} is taken:
\begin{equation}\label{eq:first-ort}
    \boldsymbol{D}^i \to
    \boldsymbol{D}^i
    -
    \sum_{j < i}
    \left<
    \phi^j
    |
    \xi^i 
    \right>
    \boldsymbol{A}^j. 
\end{equation}
Eqs.~\eqref{eq:linemin}-\eqref{eq:linemin-grad-norm} and \eqref{eq:first-ort} ensure the condition defined by Eq.~(\ref{eq:ort-constr}) holds throughout the optimization.
These steps are naturally expanded for the PCG algorithm, which requires an extra orthogonalization step.
This technique is analogous to the approach used in \textit{ab initio} calculations to compute multiple electronic bands within the self-consistent Kohn-Sham (KS) framework, where orthogonality is imposed between solutions~\cite{payne_iterative_1992}.

However, we note two considerations specific to the variational problem of polaron formation.
First, when optimizing the variational expression for a polaronic state $\boldsymbol{A}^i$, the set of previously obtained solutions must include all primitive translations within the supercell:  
\begin{equation}  
    \left\{ A^j_{n\mathbf{k}} \right\} \rightarrow \left\{ e^{i\mathbf{k \cdot R_p}} A^j_{n\mathbf{k}} \right\},~j \ne i.  
\end{equation}
Second, polaronic wavefunctions are not inherently orthogonal as their deformation field is not the same.  
Thus, enforcing the orthogonality condition may lead the process to a constraint-induced stationary point, when the minimum value of $E_\mathrm{pol}$ is reached under the constraints imposed by Eq.~\eqref{eq:ort-constr}, but the $\nabla_{A_{n\mathbf{k}}}  E_{\rm pol} \equiv 0$ condition is not fulfilled.
Consequently, the orthogonalization procedure does not necessarily lead to a local minimum but ensures that an $i$-th polaronic configuration is sufficiently distinct from prior solutions.  
Once such a configuration is found, the orthogonality constraint is removed to allow the system to converge to its local minima.
A flowchart illustrating the VarPEq self-consistent optimization process for multiple polaronic states is shown in Fig.~\ref{fig:ort_flow}.

\subsection{Long-range effects}
\label{sec:lr}

\begin{figure}[t]
    \centering
    \includegraphics[width=1.\linewidth]{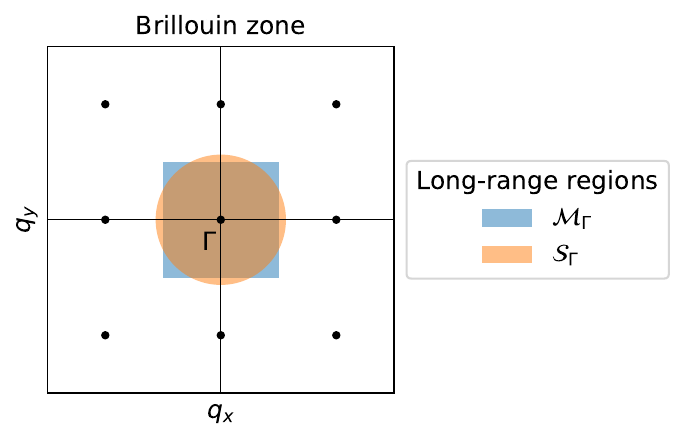}
    \caption{
    Schematic representation of BZ regions relevant to the long-range electron-phonon interaction at finite $\mathbf{q}$-mesh resolution.
    Black-filled circles denote uniform $\mathbf{q}$-mesh sampling.
    The blue-shaded square represents the microzone $\mathcal{M}_\Gamma$, capturing the missing long-range contribution to $E_\mathrm{e\text{-} ph}$.
    The orange-shaded circular region corresponds to the integration sphere \(\mathcal{S}_\Gamma\) used in numerical calculations to approximate $E_\mathrm{e\text{-} ph}$ within $\mathcal{M}_\Gamma$.
    }
    \label{fig:microzone}
\end{figure}

When optimizing VarPEq for a polaronic solution, it is essential to incorporate all contributions to the binding energy, as defined by Eqs.~\eqref{eq:varpeq}.
Here, we address the missing long-range component of the electron-phonon term from finite $\mathbf{k/q}$-sampling~\cite{Ponce2015} and present a method to account for it.
This effect is particularly significant for the EPC in polar materials, where $g_{mn\nu}(\mathbf{k}, \mathbf{q})$ exhibit a divergence in the long-wavelength limit $\mathbf{q} \rightarrow \Gamma$.
In fact, a large proportion of the EPC that influences the formation of polarons originates in the region near $\mathbf{q} = \Gamma$, which we refer to as the microzone $\mathcal{M}_\Gamma$, illustrated in Fig.~\ref{fig:microzone}.

To address this, an analytic effective long-range correction $g^\mathrm{eff}_\nu$ is added to the diagonal elements of Eq.~\eqref{eq:varpeq:elph} at $\mathbf{q} = \Gamma$:
\begin{equation}\label{eq:g_corr}
    g_{mn\nu}\left( \mathbf{k}, 0 \right)
    \rightarrow
    g_{mn\nu}\left( \mathbf{k}, 0 \right) 
    +
    \delta_{mn} g^\mathrm{eff}_\nu
\end{equation}
with
\begin{equation}\label{eq:g_eff}
    g^\mathrm{eff}_\nu
    =
    \left(\frac{1}{2}N_p
    \omega_{\Gamma\nu}
    \left\langle
    E_\mathrm{e\text{-}ph}^\mathcal{L}
    \right\rangle_{\mathcal{S}_\Gamma, \nu}
    \right)^{1/2},
\end{equation}
where $g^\mathrm{eff}_\nu$ is derived by averaging the long-range electron-phonon energy contribution over the sphere $\mathcal{S}_\Gamma$, which approximates the microzone $\mathcal{M}_\Gamma$:
\begin{multline}\label{eq:eph_avg}
    \left\langle 
    E_\mathrm{e \text{-} ph}^\mathcal{L}
    \right\rangle_{\mathcal{S}_\Gamma, \nu}
    =
    \\
    \frac{4}{\Omega_0}
    \left(
    \frac{3}{4 \pi N_p \Omega_0}
    \right)^{1/3}
    \int
    d\hat{\mathbf{q}}
    \left(
    \frac{\hat{\mathbf{q}} \cdot \mathbf{p}_\nu ( \hat{\mathbf{q}} ) }
    {
    \omega_{\hat{\mathbf{q}}\nu}
    \epsilon^\infty (\hat{\mathbf{q}})
    }
    \right)^2.
\end{multline}
Here, $\Omega_0$ is the volume of the unit cell, and the integral accounts for the angular dependence of the phonon modes.
The phonon frequencies $\omega_{\hat{\mathbf{q}}\nu}$ and mode polarities $\mathbf{p}_\nu ( \hat{\mathbf{q}} )$ depend only on the direction vector $\hat{\mathbf{q}}$ at $\Gamma$.
The notation
\begin{equation}
\epsilon^\infty(\hat{\mathbf{q}})
=
\hat{\mathbf{q}}
\cdot
\boldsymbol{\epsilon}^\infty
\cdot
\hat{\mathbf{q}}
\end{equation}
denotes the angular component of the dielectric tensor.
Additional details are provided in Appendix~\ref{appendix:g0}.
In general, the effective correction $g^\mathrm{eff}_\nu$ improves the convergence rate with the size of the supercell by modifying the precondioner in Eq.~\eqref{eq:pcond} to account for long-range effects, so it becomes
\begin{multline}\label{eq:lr-pcond}
    P^\mathcal{L}_{n\mathbf{k},n'\mathbf{k}'} \left( \varepsilon_{\rm loc} \right)
    = \\
    \delta_{n\mathbf{k},n'\mathbf{k}'}
    \left|
    \varepsilon_{n\mathbf{k}} + |\varepsilon_{\rm loc}|
    - 2
    \sum_{\nu}
    \left\langle 
    E_\mathrm{e \text{-} ph}^\mathcal{L}
    \right\rangle_{\mathcal{S}_\Gamma, \nu}
    \right|^{-1}.
\end{multline}


\subsection{Polaron size}
\label{sec:aloc}

The definition of the size of a polaron is by no means unique. 
For example, one can rely on the standard deviation $\sigma$ of the polaron density $\rho(\mathbf{r})$
\begin{equation}\label{eq:stdev}
    \sigma^2
    =
    \int d\mathbf{r}
    \left (\mathbf{r} - \mathbf{r}_c \right)^2 \rho(\mathbf{r}),
\end{equation}
where
\begin{equation}\label{eq:polaron_center}
    \mathbf{r}_c
    =
    \int d\mathbf{r}
    \,\,
    \mathbf{r} \rho(\mathbf{r})
\end{equation}
denotes the cartesian coordinates of the polaron center.

Links can then be established with well-known models like the Gaussian ansatz for the strong coupling limit~\cite{Landau1948, guster_frohlich_2021}.
For such model in the isotropic case, 
the wavefunction is given by
\begin{equation}
    \phi(\mathbf{r}) = \left(  \frac{1}{a_{\rm pol} \sqrt{\pi}} \right)^3 e^{\textstyle \frac{-(\mathbf{r} -\mathbf{r}_c)^2}{2a_{\rm pol}^2}},
\end{equation}
and the polaron radius $a_{\rm pol}$
is such that
\begin{equation}
    a_{\rm pol} = \left( \frac{2}{3} \right)^{1/2} \sigma.
\end{equation}
This will be our definition of the polaron radius.
Although the polaronic wavefunction and density are reasonably well represented by an isotropic Gaussian only in specific scenarios, this definition still gives a decent qualitative estimation for $a_{\rm pol}$ in many cases.
Moreover, it can be modified to account for the anisotropy of the polaronic density by switching to a direction-resolved integration in Eq.~\eqref{eq:stdev}.

Following Ref.~\citenum{de_melo_high-throughput_2023} that uses the same definition for the polaron radius, we choose an \textit{ad hoc} representative value of $a_{\rm pol}=10$~Bohr as the boundary to label polarons as being small or large:
\begin{equation}
    \begin{cases}
        a_{\rm pol} < 10~\text{Bohr}, & \text{(small polaron)} \\
        a_{\rm pol} \ge 10~\text{Bohr}, & \text{(large polaron)}. 
    \end{cases}
\end{equation}
Although such a separation blends the borderline medium-size polarons into the other two cases, it still provides a good classification on the basis of the polaron size.
For example, it correlates with the applicability of Fr\"ohlich-type models~\cite{de_melo_high-throughput_2023}, valid for large polarons.

\subsection{Large polarons}
\label{sec:largepol}

Polarons with large spatial extent in real space require dense $\mathbf{k/q}$-meshes to represent a supercell hosting the quasiparticle.
The direct treatment of such cases can be computationally demanding.
However, this limitation can be bypassed by reducing the electronic and vibrational subspaces only to the relevant states, which contributes to the polaron formation.

In systems with large polarons, the dominant contribution is the result of long-range interactions.
This implies that the most significant electronic states are concentrated near the band edges.
Consequently, the electronic variational coefficients $A_{n\mathbf{k}}$ decay rapidly with distance $\mathbf{k}_0 + \mathbf{q}$ from the wavevector $\mathbf{k}_0$ corresponding to the band edge $\varepsilon_0$ (VBM for hole polarons and CBM for electron polarons).

As a result, one does not need to treat the entire variational subspace, but only a subset concentrated within an energy window $\Delta \varepsilon$ around the band edge.
Numerically, this allows for a reduced $\mathbf{k}$-sampling restricted to the energy-dependent regions:
\begin{equation}
    \!\!\mathds{K}_{\Delta \varepsilon}
    \!=\!
    \begin{cases}
    \left\{
    \mathbf{k}~|~\varepsilon_0 \!-\! \varepsilon_{n\mathbf{k}} < \Delta\varepsilon
    \right\}, 
    \text{(hole polaron)}
    \\
    \left\{
    \mathbf{k}~|~\varepsilon_{n\mathbf{k}} \!-\! \varepsilon_0 < \Delta\varepsilon
    \right\}, 
    \text{(electron polaron)}.
    \end{cases}
\end{equation}
Additionally, the $\mathbf{q}$-space is restricted to include only transitions within $\mathds{K}_{\Delta \varepsilon}$.
The variational nature of the formalism ensures that by increasing the filtering range $\Delta\varepsilon$ one converges to the unfiltered solution.
From a computational perspective, energy filtering can significantly reduce costs when dealing with large polarons, and we demonstrate the effectiveness of this method in Secs.~\ref{sec:large0} and \ref{sec:large1}.
 
Note that for large polarons, the strong-coupling hypothesis is usually not valid, i.e. the self-trapping energy is negligible with respect to the one already obtained with the delocalized solution, and the weak-coupling characteristics dominate the physics of such polarons.
See Ref.~\onlinecite{de_melo_high-throughput_2023} for a study of the interplay between weak/strong coupling and large/small polarons for a large number of materials. 
Nevertheless, in the present context, even for such cases it is interesting to look at the degeneracy and symmetry breaking, even if they do not dominate the physics of such polarons.

\section{Computational details}
\label{sec:setup}

In this work, we investigate the formation of polarons in LiF, MgO, zinc-blende (zb) GaN, wurtzite (wz) GaN, and Li$_2$O$_2$.
Table~\ref{tab:systems} reports the crystal symmetry of these systems and the corresponding polaron types explored in this study.
Polarons are obtained using the VarPEq method, as recently implemented in the \textsc{ABINIT} software package~\cite{gonze_abinitproject_2020} and released in v10.4~\cite{abinit-2025}.
Variational expression is optimized via the PCG algorithm, which accounts for multiple polaronic states, long-range electron-phonon interactions, and energy filtering for large polarons, as described in Sections~\ref{sec:multipol}-\ref{sec:largepol}.
DFT calculations are performed with \textsc{ABINIT}, using PBE~\cite{Perdew1996} and PBEsol~\cite{perdew_pbesol} for the exchange-correlation (XC) energy with norm-conserving pseudopotentials~\cite{Hamann2013} provided by the \textsc{PseudoDojo} library~\cite{vanSetten2018}. 
The latest available pseudopotential versions are used: 0.5 for PBE and 0.4.1 for PBEsol. 
Table~S1 of Supplementary Information~\cite{SI} reports the relaxed structural parameters obtained with both functionals, the plane-wave energy cutoffs and the $\mathbf{k/q}$-meshes used to compute the electronic and phonon parameters for each system. 
The atomic positions and lattice parameters of each system are relaxed until the maximum absolute force falls below $10^{-6}$~Ha/Bohr.
The KS electronic bands $\varepsilon_{n\mathbf{k}}$ and wavefunctions $\psi_{n\mathbf{k}}$ are computed self-consistently on fixed $\mathbf{k}$-meshes. 
The electronic density from these calculations is then used in subsequent non-self-consistent runs to obtain these parameters on $\mathbf{k}$-meshes of arbitrary size. 
Phonon dispersions $\omega_{\mathbf{q}\nu}$, eigenmodes $e_{\kappa\alpha\nu}(\mathbf{q})$, and first-order derivatives of the KS potential $V^\mathrm{KS}$,
\begin{equation}
    \partial_{\kappa\alpha\mathbf{q}} v^\mathrm{KS}
    =\sum_p e^{-\mathbf{q} \cdot (\mathbf{r} - \mathbf{R}_p)} \left.
    \frac{\partial V^\mathrm{KS}}{\partial \tau_{\kappa\alpha}}
    \right|_{\mathbf{r} - \mathbf{R}_p},
\end{equation}
are first computed on coarse $\mathbf{q}$-meshes and subsequently interpolated onto $\mathbf{q}$-meshes of arbitrary density~\cite{brunin_phonon-limited_2020}.

The electron-phonon matrix elements
\begin{multline}
g_{mn\nu}(\mathbf{k, q}) = \sum_{\kappa\alpha}
\left(\frac{1}{2 M_\kappa \omega_{\mathbf{q}\nu}}
\right)^{1/2} e_{\kappa\alpha\nu}(\mathbf{q})
\\
\times \left\langle \psi_{m\mathbf{k+q}} | e^{i\mathbf{q} \cdot \mathbf{r}}
\partial_{\kappa\alpha\mathbf{q}} v^\mathrm{KS} |\psi_{n\mathbf{k}}
\right\rangle
\end{multline}
are computed on arbitrary $\mathbf{k/q}$-meshes by Fourier-interpolating the short-range part of the scattering potentials, while the dipole- and quadrupole-induced long-range contributions to the potentials are treated analytically~\cite{brunin_electron-phonon_2020, brunin_phonon-limited_2020}.
The dielectric tensor $\boldsymbol{\epsilon}^\infty$, Born effective charges $\boldsymbol{Z}^*$, and dynamical quadrupoles $\boldsymbol{Q}$ required for long-range treatment are computed using denser $\mathbf{k}$-meshes compared to ground-state calculations to ensure convergence.
These meshes and calculated tensor components are reported in Tables~S2-S12 of Supplementary Information~\cite{SI}.
For quadrupole calculations, pseudopotential without non-linear core corrections are used~\cite{pseudodojo_nocc_github_repo}.
Quadrupole calculations with nonlinear core corrections can be performed with the recent ABINIT v10.4 release~\cite{abinit-2025}, but this functionality was not available when most of the present work was done.

The electronic bands $\varepsilon_{n\mathbf{k}}$ and the phonon dispersions $\omega_{\mathbf{q}\nu}$ are computed within the irreducible BZ.
However, electron-phonon matrix elements $g_{mn\nu}(\mathbf{k, q})$ are evaluated in the full BZ, with both the $\mathbf{k}$- and $\mathbf{q}$-mesh symmetries disabled. 
This approach is necessary because the VarPEq formalism is gauge-dependent in $g_{mn\nu}(\mathbf{k, q})$ and must correctly describe transitions involving degenerate electronic states or phonons. 
Although symmetry unfolding and gauge recovery techniques~\cite{Li2024} could allow the use of symmetry for electron-phonon matrix elements, they are not yet implemented.
Additionally, to capture the symmetry breaking effects associated with polaron formation~\cite{vasilchenko_polarons_cubic_2024},
no symmetry constraints are imposed on the variational coefficients $\boldsymbol{A}$ and $\boldsymbol{B}$.

Post-processing and visualization of the computed data are performed using the \textsc{AbiPy} toolkit~\cite{gonze_abinitproject_2020}, which provides a comprehensive set of tools to analyze the output of \textsc{ABINIT}, including polaron calculations.

\begin{table}
\renewcommand\arraystretch{1.5}
\centering
\caption{
The set of materials studied in this work with crystal symmetry, optimized lattice constants and polaron radii $a_{\rm pol}$ (PBE XC).
Values of $a_{\rm pol} \ge 10$~Bohr corresponding to large polarons are highlighted in bold.
The $h^+$ and $e^-$ symbols indicate hole and electron polarons, respectively.
In GaN, we do not find localized $e^-$ polarons, and $a_{\rm pol}$ values correspond to the linear dimensions of the largest supercells employed in the calculations.
}
\label{tab:systems}
\begin{tabular}{cc|cc|cc}
\hline \hline
Materials   & Space group  & \multicolumn{2}{c|}{Lattice (Bohr)} &
\multicolumn{2}{c}{$a_{\rm pol}$~(Bohr)} \\
\hline
\multicolumn{2}{c|}{Cubic} & \multicolumn{2}{c|}{$a$ }  & $h^+$ & $e^-$ \\
LiF         & Fm$\overline{3}$m & \multicolumn{2}{c|}{7.6753}   & 2.26 & {\bf 11.81}  \\
MgO         & Fm$\overline{3}$m & \multicolumn{2}{c|}{8.0368}   & 4.67 & {\bf 82.33} \\
zb-GaN      & F$\overline{4}$3m & \multicolumn{2}{c|}{8.5981}   & {\bf 104.81} & $>\!\!120$  \\
\hline
\multicolumn{2}{c|}{Hexagonal} & $a$       & $c$ \\
wz-GaN      & P6$_3$mc  &  6.0842    & 9.9111  & {\bf 110.54} & $>\!\!120$  \\
Li$_2$O$_2$ & P6$_3$/mmc & 5.9695    & 14.5184 & 3.50 & 3.17 \\
\hline \hline
\end{tabular}
\end{table}

\section{Results}
\label{sec:results}

In this section, we compute the polaron formation energy in the five materials listed in Sec.~\ref{sec:setup}.
We first present results on small polarons and subsequently discuss large polarons.
These two regimes, dominated by short- and long-range electron-phonon interactions, provide qualitatively different insights into the physics of polaron formation and the applicability of the variational method.
The computed values of polaron radii $a_{\rm pol}$ that we use to distinguish between these regimes are given in Table~\ref{tab:systems} and Table~S1 of Supplementary Information~\cite{SI} for PBE and PBEsol parametrization, respectively.

\subsection{Small polarons}
\label{sec:small0}

\begin{figure*}[t]
    \centering
    \includegraphics[width=1.\linewidth]{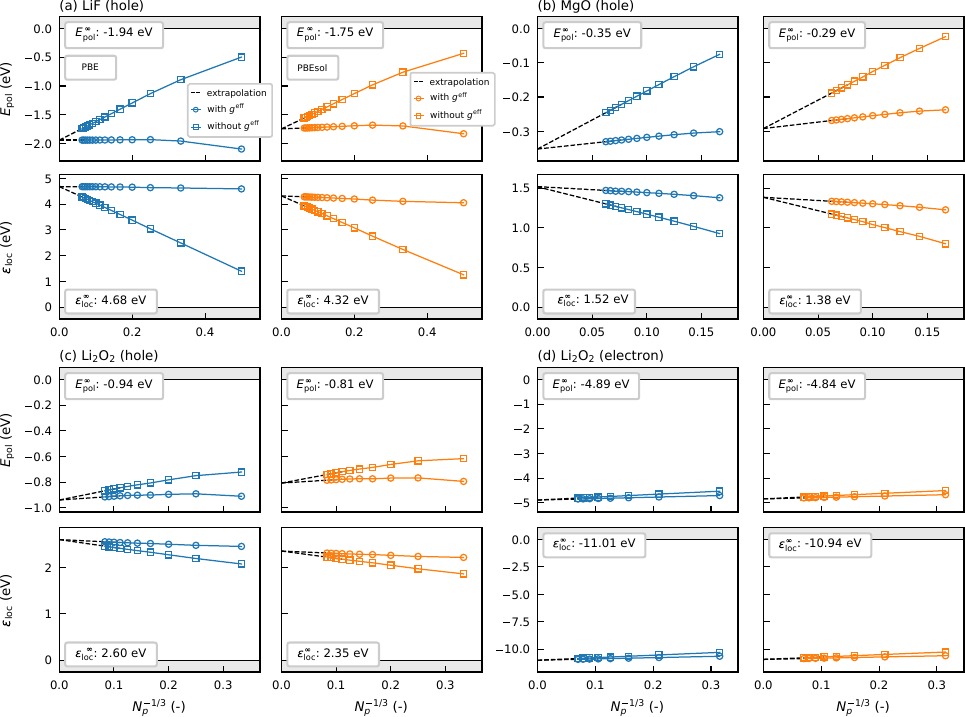}
    \caption{
        Convergence of the binding energy $E_\mathrm{pol}$ and charge localization energy $\varepsilon_{\rm loc}$ for small polarons as a function of inverse linear supercell size $N_p^{-1/3}$.
        The results correspond to:
        (a) hole in LiF,
        (b) hole in MgO,
        (c) hole and (d) electron in Li$_2$O$_2$.
        The colors indicate XC functionals : PBE in blue and PBEsol in orange.
        Open circles and squares denote calculations with and without long-range electron-phonon corrections, respectively, at fixed $\mathbf{k/q}$-point meshes.
        Dashed black lines show extrapolations to the infinite-size limit.
        Extrapolated values for $E_\mathrm{pol}^{\infty}$ and $\varepsilon_{\rm loc}^{\infty}$ are indicated within each panel.
        Regions indicating unbound polarons are shaded in gray.
        Data points are computed on $\mathbf{k/q}$-meshes of incremental density:
        (a) 2$\times$2$\times$2 $\rightarrow$ 16$\times$16$\times$16 (LiF hole),
        (b) 6$\times$6$\times$6 $\rightarrow$ 16$\times$16$\times$16 (MgO hole),
        (c) 3$\times$3$\times$3 $\rightarrow$ 12$\times$12$\times$12 (Li$_2$O$_2$ hole),
        (d) 4$\times$4$\times$2 $\rightarrow$ 18$\times$18$\times$9 (Li$_2$O$_2$ electron).
    }
    \label{fig:energy_small}
\end{figure*}

We find small polarons when adding a hole in LiF and MgO, and for the addition of an electron and a hole in the case of Li$_2$O$_2$. 
In these systems, variational optimization yields negative binding energies even at coarse $\mathbf{k/q}$-meshes: 2$\times$2$\times$2 (LiF hole), 6$\times$6$\times$6 (MgO hole), 3$\times$3$\times$3 (Li$_2$O$_2$ hole) and 4$\times$4$\times$2  (Li$_2$O$_2$ electron), indicating small polaron formation in the corresponding supercells.
To obtain accurate energies at the infinite-size supercell limit $N_p \rightarrow \infty$, we extrapolate results from progressively denser meshes using a linear extrapolation with respect to $N_p^{-1/3}$:
\begin{equation}
    E_\mathrm{pol}(N_p) = E^{\infty}_{\mathrm{pol}} + \alpha N_p^{-1/3} + \mathcal{O}(N_p^{-1}),
\end{equation}
where $\alpha$ is the fitting parameter. 
Figure~\ref{fig:energy_small} shows extrapolated binding energy $E_\mathrm{pol}$ and localization energy $\varepsilon_{\rm loc}$ computed using PBE and PBEsol parametrization.
Applying the long-range correction described in Sec.~\ref{sec:lr} brings the solutions closer to extrapolated values at finite mesh sizes.
This does not affect the extrapolated values, as the corrections vanish with the inverse supercell size as $N_p^{-1/3} \rightarrow 0$.

\begin{table}[t]
\renewcommand\arraystretch{1.5}
\centering
\caption{
Binding energies $E_\mathrm{pol}$ and charge localization energies $\varepsilon_{\rm loc}$ of small polarons in LiF, MgO, and Li$_2$O$_2$.
Results obtained in this work (PBE parametrization) are compared with data from the literature.
The $h^+$ and $e^-$ symbols indicate hole and electron polarons, respectively.
}
\label{tab:small-energy-ref}
\begin{tabular}{c|cc|cc}
  \hline \hline
                         & \multicolumn{2}{c|}{$E^\mathrm{pol}$~(eV)}
                         & \multicolumn{2}{c}{$\varepsilon^{\rm loc}$~(eV)} \\
                         & This work & References  & This work  & References   \\
  \hline
  LiF ($h^+$)             & -1.94  & -1.98 \cite{sio_ab_2019} &   4.68  &   4.76 \cite{sio_ab_2019} \\
  MgO ($h^+$)             & -0.35  & -0.50 \cite{falletta_polarons_2022} &   1.52  &  1.96 \cite{falletta_polarons_2022} \\
  Li$_2$O$_2$ ($h^+$)     & -0.94  & -0.95 \cite{Radin2013}  &   2.60  &     1.28 \cite{Radin2013} \\
  Li$_2$O$_2$ ($e^-$) & -4.89  & -4.89 \cite{sio_ab_2019} & -11.01  & -10.98 \cite{sio_ab_2019} \\
  \hline \hline
\end{tabular}
\end{table}

Computed small polarons are strongly bound, with the binding energy ranging from $E_\mathrm{pol} = -0.29/-0.35$~eV (MgO hole) to $E_\mathrm{pol} = -4.84/-4.89$~eV (Li$_2$O$_2$ electron) for PBE/PBEsol functionals, respectively.
Table~\ref{tab:small-energy-ref} compares our extrapolated PBE values with previous studies.
Our findings closely match results by Sio \textit{et al.}~\cite{sio_ab_2019,sio_polarons_2019} for hole polarons in LiF and electron polarons in Li$_2$O$_2$.
Minor differences ($\sim$2\%) arise mainly from differing approaches to interpolate electron-phonon matrix elements $g_{mn\nu}(\mathbf{k,q})$: our method uses interpolation of the perturbed potentials $\partial_{\kappa\alpha\mathbf{q}} v^\mathrm{KS}$ while Refs.~\citenum{sio_ab_2019,sio_polarons_2019} use Wannier interpolation via \textsc{EPW}~\cite{ponce_epw_2016,Lee2023}.
We validate that when no interpolation is performed, the results are identical between both approaches.

Comparisons for holes in MgO and Li$_2$O$_2$ with real-space DFT studies~\cite{falletta_many-body_2022,falletta_polarons_2022,Radin2013}
which use self-interaction corrected and tuned hybrid functionals for polaron localization, show excellent agreement for Li$_2$O$_2$ but a 30\% difference for the less bound MgO polaron.
The observed discrepancy in MgO could be explained by the fact that we consider responses limited to conduction/valence states for electron/hole polarons, linear electron-phonon coupling, and neglect the Debye-Waller self-energy~\cite{Ponce2014,Ponce2025}.
To investigate such systems, a more sophisticated approach~\cite{lafuente-bartolome_ab_2022, lafuente-bartolome_unified_2022} that takes into account dynamical polaronic effects~\cite{Stefanucci2025} may be required.
In contrast, direct DFT methods fully solve the KS Hamiltonian with an added charge, potentially more suitable for systems like MgO.

The electron polaron in Li$_2$O$_2$ highlights the importance of both conduction and valence manifolds in polaron formation.
Computed from conduction bands only, this polaron exhibits an excessively large charge localization energy $\varepsilon_\mathrm{loc}$ of 11~eV.
This value exceeds the band gap of Li$_2$O$_2$ (7.76~eV via self-consistent GW~\cite{Radin2013}).
As $\varepsilon_\mathrm{loc}$ represents the energy of the localized polaronic state, a value $\varepsilon_\mathrm{loc}$ of 11~eV artificially positions this state below the Fermi level.
Modifying the formalism to include the response from conduction and valence states may prevent this overestimation.

Finally, comparing PBE and PBEsol parametrizations reveals less bound polarons with systematically higher $E_\mathrm{pol}$ and $\varepsilon_\mathrm{loc}$ values for PBEsol.
Such differences arise primarily due to variations in the bandwidth of conduction/valence states and phonon dispersions.

\subsection{Symmetry-breaking of small polarons and degenerate solutions}
\label{sec:small_symbreak}

\begin{figure*}[t]
    \centering
    \includegraphics[width=1.\linewidth]{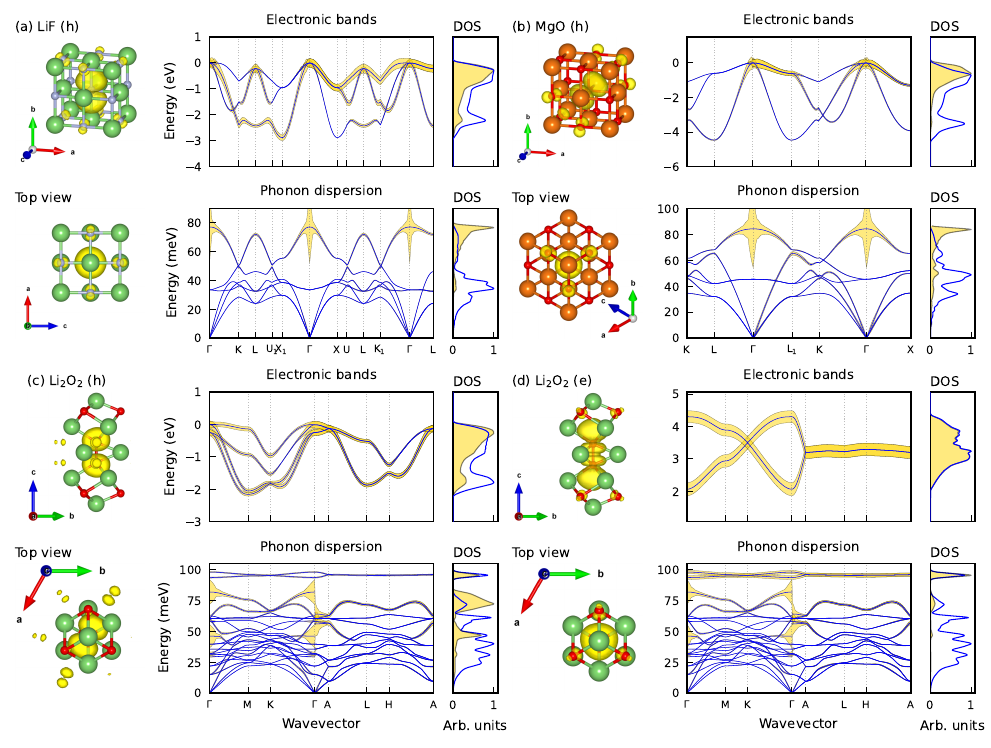}
    \caption{
    Small polaron charge densities, electron and vibrational spectral decompositions where 
    (a) hole in LiF,
    (b) hole in MgO ($D_{3d}$ symmetry),
    (d) hole in Li$_2$O$_2$,
    (f) electron in Li$_2$O$_2$.
    Polarons are obtained with PBE and $g^\mathrm{eff}$ corrections.
    Charge densities are confined to the smallest hosting unit of atoms, retaining the crystal symmetry.
    Green, gray, and orange atomic spheres represent Li, F, and O atoms, respectively.
    For each charge density, we show an angle view and a top view from the perspective of a polaron principal axis.
    The density isosurface is set at $\rho(\mathbf{r}) = 10^{-3}$~\AA$^{-3}$, except for MgO, where it is $\rho(\mathbf{r}) = 2\cdot10^{-3}$~\AA$^{-3}$.
    Ground-state electron and phonon spectra, and their corresponding DOS are shown with blue lines.
    The contribution of each component is represented by the width of a corresponding shaded gold region.
    The width of this region is proportional to $\sqrt{|A_{n\mathbf{k}}|^2_\mathrm{avg}}$ and $\sqrt{|B_{\mathbf{q}\nu}|^2_\mathrm{avg}}$ for electrons and phonons.
    %
    %
    All DOS are normalized over the displayed energy region and shown in arbitrary units.
    }
    \label{fig:bands_small}
\end{figure*}

\begin{figure}[t]
    \centering
    \includegraphics[width=1.\linewidth]{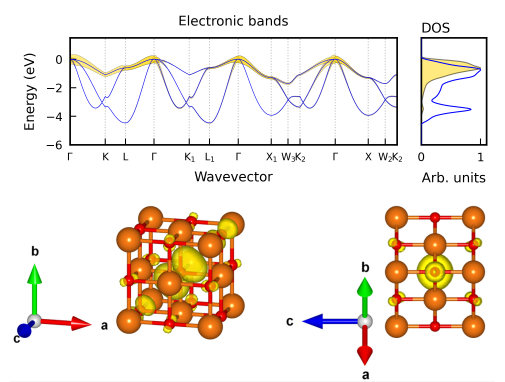}
    \caption{
    Polaron charge density and electron and vibrational spectral decompositions for the $D_{2h}$ polaron in MgO.
    The density isosurface is set at $\rho(\mathbf{r}) = 2 \cdot 10^{-3}$~\AA$^{-3}$.
    We follow the same convention as Fig.~\ref{fig:bands_small}.
    }
    \label{fig:mgo_d2h}
\end{figure}

\begin{figure}[t]
    \centering
    \includegraphics[width=1.\linewidth]{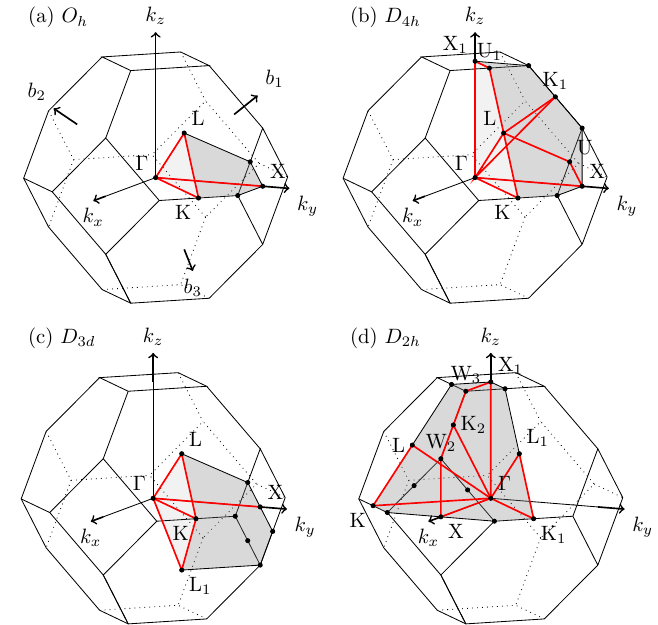}
    \caption{
    First Brillouin zone of a face-centered cubic lattice with irreducible wedges and high-symmetry paths, adjusted for symmetry breaking.
    Panel (a) shows the standard irreducible BZ with $O_h$ point group.
    The broken-symmetry irreducible BZ is shown for the point groups:
    (b) $D_{4h}$ with $\mathrm{\Gamma\text{--}X}$ principal direction;
    (c) $D_{3d}$ with  $\mathrm{\Gamma\text{--}L}$ principal direction;
    (d) $D_{2h}$ with $\mathrm{\Gamma\text{--}K}$ principal direction.
    Irreducible wedges are shaded in gray.
    High-symmetry points are shown with filled black circles, and labeled if they belong to a high symmetry-path.
    High-symmetry paths are shown with thick red lines.
    }
    \label{fig:fcc_kpath}
\end{figure}

The calculated small polarons are analyzed in real space by reconstructing their electronic wavefunctions $\phi(\mathbf{r})$ in the basis of KS states, as defined by Eq.~\eqref{eq:varphi}.
The charge localization of self-trapped polarons is represented by the polaronic density $\rho(\mathbf{r})$, given by Eq.~\eqref{eq:rho}.

In reciprocal space, we analyze the contribution of the individual electronic bands $\varepsilon_{n\mathbf{k}}$ and phonon modes $\omega_{\mathbf{q}\nu}$ to the polaron via the associated variational coefficients.
For each wavevector and state index, its contribution is proportional to $|A_{n\mathbf{k}}|^2$ and $|B_{\mathbf{q}\nu}|^2$, respectively.
For degenerate electronic and phonon states, we average such contribution as
\begin{align}
    |A_{n\mathbf{k}}|^2_\mathrm{avg}   =& \frac{1}{|n_\mathrm{deg}|} \sum_{n' \in n_\mathrm{deg}} |A_{n'\mathbf{k}}|^2 \\
    |B_{\mathbf{q}\nu}|^2_\mathrm{avg} =& \frac{1}{|\nu_\mathrm{deg}|} \sum_{\nu' \in \nu_\mathrm{deg}} |B_{\mathbf{q}\nu'}|^2,
\end{align}
where band/mode indices runs over the degenerate subspace $n_\mathrm{deg}/\nu_\mathrm{deg}$. 
The resulting average is invariant with respect to unitary transformations.

To evaluate the contribution from each electronic band and phonon mode, we also analyze the polaronic density of states (DOS), defined as
\begin{align}\label{eq:ados}
    A^2 \left(\varepsilon\right) =& \frac{1}{N_p} \sum_{n\mathbf{k}} |A_{n\mathbf{k}}|^2 \delta \left( \varepsilon - \varepsilon_{n\mathbf{k}} \right) \\
    B^2 \left(\omega\right)      =& \frac{1}{N_p} \sum_{\mathbf{q}\nu} |B_{\mathbf{q}\nu}|^2\delta \left( \omega - \omega_{\mathbf{q}\nu} \right),
    \label{eq:bdos}   
\end{align}
where the band and mode indices run over the range presented in Fig.~\ref{fig:bands_small}.
Each polaron exhibits symmetry breaking: the point-group symmetry of its charge density is lower than that of the crystal point group.
This leads to the degeneracy of the polaronic solutions, which is identified by the orthogonalization method outlined in Sec.~\ref{sec:multipol}.
Additionally, because of the symmetry breaking, the standard BZ high-symmetry paths become insufficient to analyze the electron and phonon contribution.
Hence, we use expanded high-symmetry paths, tailored for each case of symmetry-breaking.
Figure~\ref{fig:fcc_kpath} illustrates these paths for the face-centered cubic lattice.

For the hole polaron in LiF, we obtain a triply-degenerate polaron with $D_{4h}$ point group, smaller than the $O_h$ point group of the crystal.
The charge is localized at the fluorine atoms, and three degenerate solutions align with the cubic directions of the $\langle 100 \rangle$ family.
Figure~\ref{fig:bands_small}(a) illustrates one of the solutions, oriented along the $[001]$ direction.
The expanded high-symmetry path for this polaron is shown in Fig.~\ref{fig:fcc_kpath}(b).
As one of the $\mathrm{\Gamma\text{--}X}$ directions of the crystal is inequivalent with the other two, the electronic contributions for $\mathrm{\Gamma\text-X}$/$\mathrm{\Gamma\text{--}X_1}$ and $\mathrm{\Gamma\text{--}K}$/$\mathrm{\Gamma\text{--}K_1}$ lines are asymmetric.
Taking into account the degeneracies, the charge distributions of our polarons are in agreement with the polaron reported in Refs.~\citenum{sio_ab_2019, sio_polarons_2019}.

In MgO, which has the $O_h$ crystal point group, we find two distinct degenerate polarons with $D_{3d}$ and $D_{2h}$ point groups, both being localized on the oxygen atoms.
Figures~\ref{fig:bands_small}(b) and Fig.~\ref{fig:mgo_d2h} illustrate these solutions.
The former is four-fold degenerate and aligned with one of the $\langle 111 \rangle$ directions.
The latter is six-fold degenerate, with its principal axis being along one of the $\langle 110 \rangle$ directions.
Similarly to the LiF case, the high-symmetry paths are adapted for symmetry breaking, as shown in Fig.~\ref{fig:fcc_kpath}(c,d).
For the $D_{3d}$ polaron, the principal $\mathrm{\Gamma\text{--}L}$ direction is not equivalent to the other three, and the electronic contribution at the $\mathrm{\Gamma\text{--}L}$/$\mathrm{\Gamma\text{--}L_1}$ lines is asymmetric.
For the $D_{2h}$ polaron, the symmetry-breaking effect is demonstrated at the triplet of $\mathrm{\Gamma\text{--}K}$/$\mathrm{\Gamma\text{--}K_1}$/$\mathrm{\Gamma\text{--}K_2}$ lines.

For both $D_{3d}$ and $D_{2h}$ polarons in MgO, the extrapolated binding energy $E_\mathrm{pol}$ and the charge localization energy $\varepsilon_\mathrm{loc}$ differ negligibly ($\sim$~0.05\%), in favor of the  $D_{3d}$ solution.
The potential energy surface near the $D_{2h}$ solution is shallow, and convergence toward the solution is slower compared to the $D_{3d}$ polaron.
Given their minimal energy difference, accurately distinguishing them via SCF optimization can be challenging.
Both polarons were obtained in previous \textit{ab initio} studies: instances of $D_{2h}$ and $D_{3d}$ solutions are reported in Ref.~\citenum{kokott_first-principles_2018} and Refs.~\citenum{falletta_many-body_2022,falletta_polarons_2022}, respectively.
However, using the variational formalism of the present work and the generalized Fr\"ohlich model, the $D_{3d}$ polaron in MgO can be demonstrated to be favorable~\cite{vasilchenko_polarons_cubic_2024}, which is in agreement with the current first-principles results.

In Li$_2$O$_2$ with $D_{6h}$ crystal point group, both electron and hole polarons are located in the dimer of oxygen atoms.
The hole polaron in Li$_2$O$_2$ exhibits $C_{2v}$ symmetry, as shown in Fig.~\ref{fig:bands_small}(c).
Our PBEsol results closely resemble previously reported hole polaron in Ref.~\citenum{Radin2013}.
Similarly to the $D_{2h}$ polaron in MgO, the potential energy surface near this solution is shallow, and optimization can be challenging.
The electron polaron in Li$_2$O$_2$ has $D_{3h}$ point group, as shown in Fig.~\ref{fig:bands_small}(d), and its charge localization is consistent with that reported in Refs.~\citenum{sio_ab_2019,sio_polarons_2019}.

\subsection{Large polarons}
\label{sec:large0}

\begin{figure*}[t]
    \centering
    \includegraphics[width=1.\linewidth]{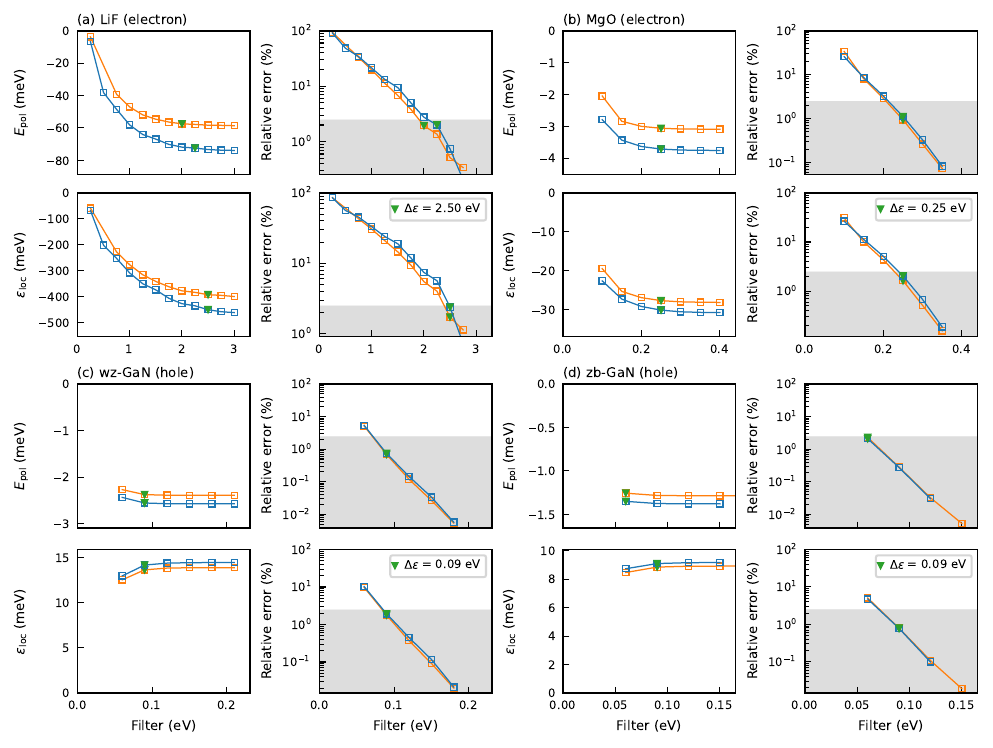}
    \caption{
    Convergence of the binding energy $E_\mathrm{pol}$ and charge localization energy $\varepsilon_\mathrm{loc}$ of large polarons as a function of the energy filtering window $\Delta \varepsilon$ for states near the band edges.
    The results are obtained at fixed $\mathbf{k/q}$-mesh and correspond to:
    (a) electron in LiF (20$\times$20$\times$20),
    (b) electron in MgO (20$\times$20$\times$20),
    (c) hole in wz-GaN  (74$\times$74$\times$37),
    (d) hole in zb-GaN  (77$\times$77$\times$77).
    The colors indicate XC functionals with PBE (blue) and PBEsol (orange).
    Open squares denote calculations without long-range electron-phonon corrections at fixed values of $\Delta \varepsilon$.
    Filled green triangles indicate the convergent values of $\Delta \varepsilon$, entering the 2~\% relative error threshold, shown by the shaded gray region.
    The upper bounds for $\Delta \varepsilon$ required for the convergence of $\varepsilon_\mathrm{loc}$ are indicated within each panel.
    }
    \label{fig:efilter}
\end{figure*}

\begin{figure*}[t]
    \centering
    \includegraphics[width=1.\linewidth]{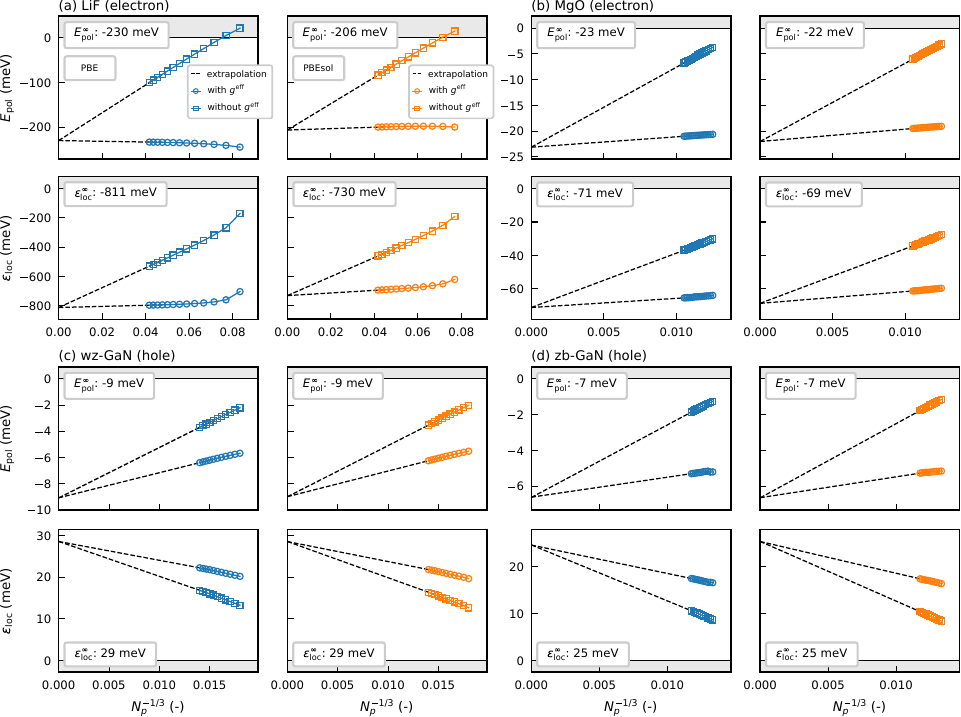}
    \caption{
        Convergence of the binding energy $E_\mathrm{pol}$ and charge localization energy $\varepsilon_{\rm loc}$ of large polarons as a function of inverse linear supercell size $N_p^{-1/3}$.
        The results correspond to
        (a) electron in LiF,
        (b) electron in MgO,
        (c) hole in wz-GaN,
        (d) hole in zb-GaN.
        In these calculations, the energy filtering technique is utilized as follows:
        (a) no filtering,
        (b) $\Delta \varepsilon = 250$~meV,
        (c-d)  $\Delta \varepsilon = 90$~meV.
        The colors indicate XC functionals with PBE (blue) and PBEsol (orange).
        Open circles and squares denote calculations with and without long-range electron-phonon corrections, respectively, at fixed $\mathbf{k/q}$-point meshes.
        Dashed black lines show extrapolations to the infinite-size limit.
        Extrapolated values for $E_\mathrm{pol}$ and $\varepsilon_{\rm loc}$ are indicated within each panel.
        Regions indicating unbound polarons are shaded in gray.
        Data points are computed on $\mathbf{k/q}$-meshes of incremental density:
        (a) 12$\times$12$\times$12 $\rightarrow$ 24$\times$24$\times$24 (LiF electron),
        (b) 80$\times$80$\times$80 $\rightarrow$ 95$\times$95$\times$95 (MgO hole),
        (c) 70$\times$70$\times$35 $\rightarrow$ 90$\times$90$\times$45 (wz-GaN hole),
        (d) 75$\times$75$\times$75 $\rightarrow$ 85$\times$85$\times$85 (zb-GaN hole).
    }
    \label{fig:energy_large}
\end{figure*}

Large electron polarons are found in LiF and MgO, and both polymorphs of GaN exhibit large hole polaron formation.
In these systems, variational optimization requires dense $\mathbf{k/q}$-meshes as the polarons are localized in reciprocal space around the band edges.
To optimize these polarons, we employ the energy-filtering technique described in Sec.~\ref{sec:lr}.

First, at fixed $\mathbf{k/q}$-mesh we determine the value of the energy filtering window $\Delta \varepsilon$, required for convergence of $E_\mathrm{pol}$ and $\varepsilon_\mathrm{loc}$.
The latter quantity systematically requires larger values of $\Delta \varepsilon$ for convergence.
Moreover, optimization without $g^\mathrm{eff}$ corrections also requires larger values of $\Delta \varepsilon$.
As a result, we select the filtering energy required to converge $\varepsilon_\mathrm{loc}$ without $g^\mathrm{eff}$ as an upper bound for $\Delta \varepsilon$:
2.5~eV for LiF, 0.25~eV for MgO, and 0.09~eV for both polymorphs of GaN.
The results of the convergence studies are summarized in Fig.~\ref{fig:efilter}.
The estimated speedup $s_{\Delta \epsilon}$ is system-dependent. It can be calculated thanks to the ratio between the number of $\mathbf{k}$-points in the full BZ and the filtered region
\begin{equation}
    s_{\Delta \epsilon} = \frac{|\mathds{K}_{\rm BZ}|}{|\mathds{K}_{\Delta \varepsilon}|}.
\end{equation}
The average of $s_{\Delta \epsilon}$ over various $\mathbf{k}$-meshes for selected values of $\Delta \varepsilon$ are listed in Table.~S13 of Supplementary Information~\cite{SI}.
At fixed values of $\Delta \varepsilon$, we perform the same optimization and extrapolation procedure as discussed in Sec.~\ref{sec:small0}.
The results are shown in Fig.~\ref{fig:energy_large} where the optimization of LiF is shown without filtering.

Using LiF as an example, we validate the filtering technique.
When $\Delta \varepsilon = 2.5$~eV filtering is used, the extrapolated values are $E_\mathrm{pol} = -229/206$~meV, $\varepsilon_\mathrm{loc} = -794/718$~meV for PBE/PBEsol XC.
Without filtering, these values become $E_\mathrm{pol} = -230/206$~meV, $\varepsilon_\mathrm{loc} = -811/730$~meV, validating the energy-filtered calculations.
The unfiltered results are also in agreement with the ones reported in Refs.~\citenum{sio_ab_2019,sio_polarons_2019}.

Polarons in MgO and wz/zb-GaN require dense sampling of the BZ near the band edges and are computationally prohibitive without filtering.
In these systems, curvatures of the band extrema and phonons in the vicinity of $\mathbf{q} \to \Gamma$ become the dominating factors for the polaron formation, hence the difference between the PBE and PBEsol XC are negligible since they give the same band curvature.
The extrapolated values range from $E_\mathrm{pol} = -23$~meV and $\varepsilon_\mathrm{loc} = -71$~meV for electron MgO to $E_\mathrm{pol} = -7$~meV and $\varepsilon_\mathrm{loc} = -24$~meV for hole in zb-GaN.

For MgO, with isotropic electron effective mass, $E_\mathrm{pol}$ and $\varepsilon_\mathrm{loc}$ are in agreement with the values that can be obtained with the adiabatic treatment of the standard Fr\"ohlich model in the strong-coupling regime \cite{j_miyake_strong-coupling_1975, j_miyake_ground_1976}:
\begin{align}
    E^{\rm Fr}_\mathrm{pol} & = -0.1085~\alpha^2 \omega_{\rm LO}, \\
    \varepsilon^{\rm Fr}_{\mathrm{loc}} & = 3 E_\mathrm{pol}.
\end{align}
Here,
\begin{equation}
    \alpha =
    \left(
    \frac{m^*}{2\omega_{\rm LO}}
    \right)^{1/2}
    \left(
    \epsilon^*
    \right)^{-1}
\end{equation}
is the dimensionless coupling constant, defined by the macroscopic parameters of a system: effective mass $m^*$, dispersionless LO mode frequency $\omega_{\rm LO}$, and static and high-frequency dielectric constants
\begin{equation}
(\epsilon^*)^{-1} = (\epsilon^\infty)^{-1} - (\epsilon^0)^{-1}.
\end{equation}
For PBE/PBEsol input, $E^{\rm Fr}_\mathrm{pol}$ = -24/-23~meV, consistent with the extrapolated values of $E_\mathrm{pol}$, obtained variationally.
The values of the corresponding Fr\"ohlich parameters are listed in Table~S14 of Supplementary Information~\cite{SI}.

Notably, the polarons in MgO and GaN possess binding energies smaller than the most coupled phonon frequencies in these systems.
This indicates that, provided the hosting supercell represented by the variational subspace is large enough, it is possible to achieve self-trapping of weakly-bound polarons.
However, as these quasiparticles are weakly bound, their correct treatment will require an extension of the present formalism to include effects beyond self-trapping.

\subsection{Symmetry-breaking of large polarons}
\label{sec:large1}

\begin{figure*}[t]
    \centering
    \includegraphics[width=1.\linewidth]{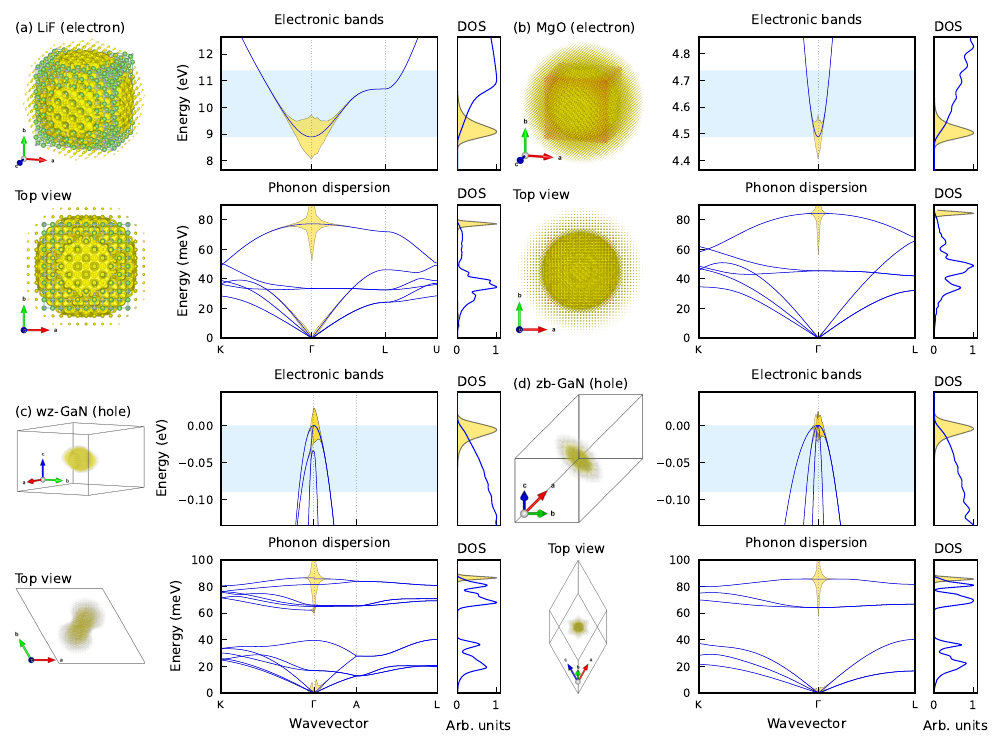}
    \caption{
    Large polaron charge densities, electron and vibrational spectral decompositions where we show in 
    (a) electron in LiF,
    (b) electron in MgO,
    (c) hole in wz-GaN,
    (d) hole in zb-GaN,
    optimized for PBE parametrization with $g^\mathrm{eff}$ corrections.
    For (a-b), charge densities are confined to a hosting unit of atoms, retaining the crystal symmetry.
    For (c-d), charge densities are shown within the primitive simulation cell.
    For each charge density, we show an angle view and a top view from the perspective of a polaron principal axis.
    The density isosurface is set at $\rho(\mathbf{r}) = 5\cdot10^{-7}$~\AA$^{-3}$.
    Ground state electron and phonon spectrum and DOS are shown with blue lines.
    The contribution of each component is represented by the width of a corresponding shaded gold region.
    Shaded blue regions near the band edges represent the filtering energy windows $\Delta \varepsilon$ used in the calculations.
    The width of this region is proportional to $\sqrt{|A_{n\mathbf{k}}|^2_\mathrm{avg}}$ and $\sqrt{|B_{\mathbf{q}\nu}|^2_\mathrm{avg}}$ for electrons and phonons.
    All DOS are normalized and shown in arbitrary units.
    }
    \label{fig:bands_large}
\end{figure*}

For calculated large polarons, we analyze the charge localization and individual electronic and phonon contributions to the polaron formation, similar to Sec.~\ref{sec:small_symbreak}.
The obtained results are shown in Fig.~\ref{fig:bands_large}.

Spectral decomposition of the electronic part of the polaron wavefunction shows that electronic states, important for the formation of polarons, are confined within the energy windows $\Delta \varepsilon$ near the band edges.
As expected, polaron formation is mediated by optical phonon modes at $\mathbf{q} \to \Gamma$.
Hence, the self-trapping of such polarons is well described by the generalized Fr\"ohlich model~\cite{miglio_predominance_2020, vasilchenko_polarons_cubic_2024}.

The charge distribution of the polaron in LiF possesses the $O_h$ symmetry of the crystal, in agreement with Refs.~\citenum{sio_ab_2019,sio_polarons_2019}. 
The polaron in MgO is spherical and can be described within the isotropic Fr\"ohlich model.
We find no symmetry breaking in these systems as only a single band participates in the polaron formation.
However, for systems with degeneracy near the band edges, we find symmetry breaking in wz-GaN with $D_{2h}$ symmetry, while zb-GaN hosts a $D_{3d}$ polaron.
Notably, the charge localization of the latter is a trigonal antiprism, oriented along one of the $\langle111\rangle$ cubic directions.
The same shape can be achieved for the variational solution of the generalized Fr\"ohlich model solved for this system~\cite{vasilchenko_polarons_cubic_2024}.
We do not further probe these systems for degenerate solutions as the construction of the manifold of translated solutions, described in Sec.~\ref{sec:scf} is computationally prohibitive at dense grids.
However, random initialization of $\boldsymbol{A}$ results in different orientations of optimized polarons with the same energy confirms the presence of degenerate solutions.

For both polymorphs of GaN, we observe a noticeable system-size dependence of $E_\mathrm{pol}$ and $\varepsilon_{\rm loc}$ even with the inclusion of $g^{\rm eff}$, in contrast to all other cases.
This behavior comes from the divergence of the deformation field $\boldsymbol{B}$, associated with LA phonons in the long-wavelength limit:
\begin{equation}
    \lim_{\mathbf{q} \to \Gamma} B_{\mathbf{q}\nu}
    \sim
    |\mathbf{q}|^{\gamma -1},
\end{equation}
which follows from Eq.~\eqref{eq:phgrad}, combined with $\omega_{\mathbf{q}\nu} \sim |\mathbf{q}|$ and $g_{mn\nu} (\mathbf{k,q}) \sim |\mathbf{q}|^\gamma$ for LA modes near $\Gamma$.
In piezoelectric materials such as GaN, $\gamma = 1/2$, resulting in the LA divergence of $\boldsymbol{B}$.
In contrast, in nonpiezoelectric systems $\gamma = 1$, hence no such divergence occurs in the other materials we investigate.
However, this divergence as well as the Fr\"ohlich $\gamma=0$ divergence for the LO modes are integrable, and the phonon-related observables ($E_{\rm ph}$, $E_{\rm e\text{-}ph}$, $\Delta \boldsymbol{\tau}$) converge to finite values in the limit of infinite size in all cases.
To completely treat the finite-size effects in piezolectric systems, the piezoelectic divergence of $\boldsymbol{B}$ would have to be included on the same footing as the Fr\"ohlich divergence, thus introducing an extra $g^{\rm eff}$ correction for LA phonons.


\section{Conclusion}
\label{conclusion}

In this work, we investigate the formation of self-trapped polarons in polar materials via the variational polaron equations framework.
The implementation, available in the ABINIT v10.4 release, includes the corrections for long-range electron-phonon interactions and can optimize multiple polaronic solutions within a single system.
We use it to study several prototypical semiconductors and insulators for small and large polaron formation and obtain localized polaronic solutions in all cases.
We find that when a system has degenerate bands, polaron formation leads to symmetry breaking and non-trivial degenerate solutions.
For small polarons, we obtain the entire set of degenerate solutions via an orthogonalization process in the variational optimization.
For cubic systems, we propose modified high-symmetry paths in order to take into account the symmetry-breaking effect of polaron formation.
We also demonstrate the computational benefits of the filtering of the electronic states near the band edges. 
The obtained solutions are in agreement with previous studies, including the ones obtained with similar approach, DFT optimization and model approaches.
Overall, we find that the polaron potential energy surface can be shallow and highly complex, possessing multiple minima which correspond to distinct polaronic solutions. 
Our work opens up the way for further generalization including the response from both conduction and valence manifold as well as the inclusion of dynamical effects, which 
may provide further insights into polaron formation.

The data that support the findings of this article are openly available on the Materials Cloud Archive~\cite{matcloud_archive}.

\begin{acknowledgments}
V.V. acknowledges funding by the FRS-FNRS Belgium through FRIA.
S. P. is a Research Associate of the Fonds de la Recherche Scientifique - FNRS.
This publication was supported by the Walloon Region in the strategic axe FRFS-WEL-T.
Computational resources have been provided by the supercomputing facilities of the Universit\'e catholique de Louvain (CISM/UCL), the Consortium des Equipements de Calcul Intensif en F\'ed\'eration Wallonie Bruxelles (CECI) funded by the FRS-FNRS under Grant No. 2.5020.11 and computational
resources on Lucia, the Tier-1 supercomputer of the Walloon Region with infrastructure funded by the Walloon Region under the Grant Agreement No. 1910247.
\end{acknowledgments}

\appendix
\renewcommand\thefigure{A\arabic{figure}}
\setcounter{figure}{0}
\setcounter{table}{0}   

\section{Long-range EPC correction}
\label{appendix:g0}

In this section, we provide a derivation for the long-range EPC correction used in the VarPEq method.
To account for the missing fraction of long-range effects, we focus on approximating the EPC at $\mathbf{q} = \Gamma$.
This is achieved by including the missing long-range contribution in the electron-phonon term $E_\mathrm{e \text{-} ph}$ to polaron formation energy Eq.~\eqref{eq:varpeq:elph}.
In $\mathbf{q}$-space, the missing contribution to $E_\mathrm{e \text{-} ph}$ is confined to the microzone $\mathcal{M}_\Gamma$, shown in Fig~\ref{fig:microzone}.
For a given $\mathbf{k/q}$-mesh, the volume of the microzone is inversely related to the size of the Born-von Karman supercell $N_p$ as
\begin{equation}\label{eq:microvol}
    \Omega_{\mathcal{M}_\Gamma} = \frac{1}{N_p} \Omega_{BZ},
\end{equation}
where $\Omega_{BZ}$ is the BZ volume.

Within the microzone, the average long-range contribution to the electron-phonon term in Eq.~\eqref{eq:varpeq:elph} is expressed as
\begin{multline}\label{eq:eph_lr}
    \left\langle 
    E^\mathcal{L}_\mathrm{e \text{-} ph}
    \right\rangle_{\mathcal{M}_\Gamma}
    \approx
    - \frac{1}{N_p} \sum_{\substack{mn \mathbf{k}}} \sum_\nu
    \\
    \times
    \frac{1}{\Omega_{\mathcal{M}_\Gamma}}
    \int_{\mathcal{M}_\Gamma} d\mathbf{q}
    A^*_{m\mathbf{\mathbf{k}}}
    {B_{\mathbf{q}\nu} ^\mathcal{L}}^*g^\mathcal{L}_{mn\nu}(\mathbf{k, q})
    A_{n\mathbf{\mathbf{k}}} + \mathrm{(c.c)},
\end{multline}
where the long-range deformation field is obtained from Eq.~\eqref{eq:phgrad} as
\begin{equation}\label{eq:phgrad_lr}
    B^\mathcal{L}_\mathbf{\mathbf{q}\nu}
    = \frac{1}{N_p}\sum_{mn\mathbf{k}}
    A^*_{m\mathbf{k}}
    \frac{g^\mathcal{L}_{mn\nu}(\mathbf{k, q})}{\omega_{\hat{\mathbf{q}}\nu}}
    A_{n\mathbf{k}}.
\end{equation}
In the expressions above, the long-range electron-phonon interaction is approximated via the Fr\"ohlich model~\cite{Verdi2015, miglio_predominance_2020}
\begin{multline}\label{eq:g_lr}
    g^\mathcal{L}_{mn\nu}(\mathbf{k}, \mathbf{q})
    =
    \frac{1}{|\mathbf{q}|}
    \frac{4\pi}{\Omega_0}
    \left(
    \frac{1}{2N_p\omega_{\hat{\mathbf{q}}\nu}}
    \right )^{1/2}
    \frac{\hat{\mathbf{q}} \cdot \mathbf{p}_\nu ( \hat{\mathbf{q}} ) }
    {
    \epsilon^\infty (\hat{\mathbf{q}})
    }
    \\
    \times
    \left\langle
    u_{m\mathbf{k+q}}
    |
    u_{n\mathbf{k}}
    \right\rangle,
\end{multline}
where $\Omega_0$ is the unit cell volume, $u_{n\mathbf{k}} (\mathbf{r})$ are the periodic components of the Bloch functions.
Phonon frequencies $\omega_{\hat{\mathbf{q}}\nu}$ and mode polarities $\mathbf{p}_\nu ( \hat{\mathbf{q}} )$ depend only on direction  vector $\hat{\mathbf{q}}$ at $\Gamma$.
The  shorthand notation
$\epsilon(\hat{\mathbf{q}})
=
\hat{\mathbf{q}}
\cdot
\boldsymbol{\epsilon}^\infty
\cdot
\hat{\mathbf{q}}$
is used for the angular component of the dielectric tensor.

To proceed, the scalar product in the definition of $g^\mathcal{L}_{mn\nu}(\mathbf{k}, \mathbf{q})$ is Taylor expanded in $\mathbf{q}$ to first order
$\left\langle u_{m\mathbf{k+q}} | u_{n\mathbf{k}} \right\rangle \approx \delta_{nm}$~\cite{Ponce2023}.
Using Eqs.~\eqref{eq:microvol}-\eqref{eq:g_lr} and applying the normalization condition from Eq.~\eqref{eq:anorm}, the average long-range electron-phonon term is simplified to the sum of mode-resolved integrals over the microzone:
\begin{multline}\label{eq:eph_lr2}
    \left\langle 
    E^\mathcal{L}_\mathrm{e \text{-} ph}
    \right\rangle_{\mathcal{M}_\Gamma}
    \approx
    \left(\frac{4\pi}{\Omega_0}\right)^2
    \sum_\nu
    \\
    \times
    \frac{1}{\Omega_{BZ}}
    \int_{\mathcal{M}_\Gamma} d\mathbf{q}
    \frac{1}{|\mathbf{q}|^2}
    \left(
    \frac{\hat{\mathbf{q}} \cdot \mathbf{p}_\nu ( \hat{\mathbf{q}} ) }
    {
    \omega_{\hat{\mathbf{q}}\nu}
    \epsilon^\infty (\hat{\mathbf{q}})
    }
    \right)^2.
\end{multline}
We then approximate the integration over the microzone $\mathcal{M}_\Gamma$ with an integration over a $\Gamma$-centered sphere $\mathcal{S}_\Gamma$ of the same volume, shown in Fig.~\ref{fig:microzone}.
This results in the following simplified expression:
\begin{equation}\label{eq:eph_lr3}
    \left\langle 
    E^\mathcal{L}_\mathrm{e \text{-} ph}
    \right\rangle_{\mathcal{M}_\Gamma}
    \approx
    \sum_\nu
    \left\langle 
    E^\mathcal{L}_\mathrm{e \text{-} ph}
    \right\rangle_{\mathcal{S}_\Gamma, \nu}
    ,
\end{equation}
where each mode-specific contribution is given by
\begin{equation}\label{eq:eph_lr3-a}
    \left\langle 
    E^\mathcal{L}_\mathrm{e \text{-} ph}
    \right\rangle_{\mathcal{S}_\Gamma, \nu}
    \approx
    \frac{2}{\pi \Omega_0}
    q_{\mathcal{S}_\Gamma}
    \int
    d\hat{\mathbf{q}}
    \left(
    \frac{\hat{\mathbf{q}} \cdot \mathbf{p}_\nu ( \hat{\mathbf{q}} ) }
    {
    \omega_{\hat{\mathbf{q}}\nu}
    \epsilon^\infty (\hat{\mathbf{q}})
    }
    \right)^2,
\end{equation}
and
\begin{equation}
    q_{\mathcal{S}_\Gamma} = 
    2\pi
    \left(
    \frac{3}{4\pi N_p \Omega_0}
    \right)^{1/3}
\end{equation}
denotes the radius of the sphere $\mathcal{S}_\Gamma$.
The average long-range electron-phonon energy terms, as given by Eq.~\eqref{eq:eph_lr3-a} can be easily computed.

Alternatively, within the VarPEq formalism, the missing long-range contribution can be incorporated via the effective long-range correction $g^\mathrm{eff}_\nu$ to the diagonal electron-phonon matrix elements at $\mathbf{q}=\Gamma$:
\begin{equation}
    g_{mn\nu}\left( \mathbf{k}, 0 \right)
    \rightarrow
    g_{mn\nu}\left( \mathbf{k}, 0 \right) 
    +
    \delta_{mn} g^\mathrm{eff}_\nu.
\end{equation}
Under this assumption, the average effective contribution over a unit cell is expressed as
\begin{equation}\label{eq:eph_lr4}
    \left\langle 
    E^\mathcal{L}_\mathrm{e \text{-} ph}
    \right\rangle_{N_p}
    \approx
    \frac{2}{N_p}
    \sum_\nu
    \frac{|g^\mathrm{eff}_{\nu}|^2}{\omega_{\Gamma\mathbf\nu}}.
\end{equation}
By associating the average long-range electron-phonon contributions given by Eqs.~(\ref{eq:eph_lr3}) and (\ref{eq:eph_lr4}), the effective correction for the matrix elements at $\mathbf{q}=\Gamma$ is obtained:
\begin{equation}\label{eq:g_eff2}
    g^\mathrm{eff}_\nu
    =
    \left(
    \frac{1}{2}
    N_p
    \omega_{\Gamma\nu}
    \left\langle
    E^\mathcal{L}_\mathrm{e \text{-} ph}
    \right\rangle_{\mathcal{S}_\Gamma, \nu}
    \right)^{1/2}.
\end{equation}

\newpage
\bibliography{main} 

\end{document}


\title{
Supplementary Information: Variational first-principles approach to self-trapped polarons}

\author{Vasilii Vasilchenko}
\affiliation{European Theoretical Spectroscopy Facility, Institute of Condensed Matter and Nanosciences, Universit\'{e} catholique de Louvain, Rue de l'observatoire 8, bte L07.03.01, B-1348 Louvain-la-Neuve, Belgium}
\author{Matteo Giantomassi}
\affiliation{European Theoretical Spectroscopy Facility, Institute of Condensed Matter and Nanosciences, Universit\'{e} catholique de Louvain, Rue de l'observatoire 8, bte L07.03.01, B-1348 Louvain-la-Neuve, Belgium}
\author{Samuel Ponc\'e}
\affiliation{European Theoretical Spectroscopy Facility, Institute of Condensed Matter and Nanosciences, Universit\'{e} catholique de Louvain, Rue de l'observatoire 8, bte L07.03.01, B-1348 Louvain-la-Neuve, Belgium}
\affiliation{WEL Research Institute, avenue Pasteur 6, 1300 Wavre, Belgium}
\author{Xavier Gonze}
\affiliation{European Theoretical Spectroscopy Facility, Institute of Condensed Matter and Nanosciences, Universit\'{e} catholique de Louvain, Rue de l'observatoire 8, bte L07.03.01, B-1348 Louvain-la-Neuve, Belgium}

\date{\today}

\maketitle

\onecolumngrid


\section{Supplementary tables}

\begin{table}[htbp]
\renewcommand\arraystretch{1.5}
\centering
\caption*{\parbox[t]{\linewidth}{\justifying 
{TABLE S1:} The set of materials studied in this work with crystal symmetry, optimized lattice constants and polaron radii $a_{\rm pol}$ (PBEsol XC).
Values of $a_{\rm pol} \ge 10$~Bohr corresponding to large polarons are highlighted in bold.
The $h^+$ and $e^-$ symbols indicate hole and electron polarons, respectively.
In GaN, we do not find localized $e^-$ polarons, and $a_{\rm pol}$ values correspond to the linear dimensions of the largest supercells employed in the calculations.
}}
\begin{tabular}{cc|cc|cc}
\hline \hline
Materials   & Space group  & \multicolumn{2}{c|}{Lattice (Bohr)} &
\multicolumn{2}{c}{$a_{\rm pol}$~(Bohr)} \\
\hline
\multicolumn{2}{c|}{Cubic} & \multicolumn{2}{c|}{$a$ }  & $h^+$ & $e^-$ \\
LiF         & Fm$\overline{3}$m & \multicolumn{2}{c|}{7.5617}   & 2.39 & {\bf 12.38}  \\
MgO         & Fm$\overline{3}$m & \multicolumn{2}{c|}{7.9635}   & 4.93 & {\bf 83.07} \\
zb-GaN      & F$\overline{4}$3m & \multicolumn{2}{c|}{8.5032}   & {\bf 105.87} & $>\!\!120$   \\
\hline
\multicolumn{2}{c|}{Hexagonal} & $a$       & $c$ \\
wz-GaN      & P6$_3$mc  &  6.0173    & 9.8003  & {\bf 111.08} & $>\!\!120$ \\
Li$_2$O$_2$ & P6$_3$/mmc & 5.8852    & 14.3587 & 3.61 & 3.18 \\
\hline \hline
\end{tabular}
\end{table}

\begin{table}[htbp]
\renewcommand\arraystretch{1.5}
\centering
\caption*{\parbox[t]{\linewidth}{\justifying 
{TABLE S2:} The set of materials studied in this work with optimized lattice constants and computational parameters of groundstate and DFPT calculations.
$E_\mathrm{cut}$ denotes the plane-wave energy cutoff.
Listed $\mathbf{k}$- and $\mathbf{q}$-meshes represent the minimal $\Gamma$-centered uniform grids required for convergence of:
(i) KS electronic bands $\varepsilon_{n\mathbf{k}}$, wavefunctions $\psi_{n\mathbf{k}}$, phonon dispersions $\omega_{\mathbf{q}\nu}$ and eigenmodes $e_{\kappa\alpha, \nu} (\mathbf{q})$;
(ii) dielectric tensor $\boldsymbol{\epsilon}^\infty$ and Born effective charges $\boldsymbol{Z}$;
(iii) quadrupoles $\boldsymbol{Q}$.
For cubic systems with the Fm$\overline{3}$m space group, $\boldsymbol{Q} \equiv 0$ by symmetry, and aren't explicitly computed.
}}
\label{tab:systems}
\begin{tabular}{c|c|cc|cc|c|c|c|c}
\hline \hline
Materials   & Space group       & \multicolumn{4}{c|}{Lattice (Bohr)} & $E_{\rm cut}$ (Ha)  & \multicolumn{3}{c}
{$\mathbf{k/q}$-mesh} \\
\hline
            &                   & \multicolumn{2}{c|}{PBE}     & \multicolumn{2}{c|}{PBEsol} &
            &  $\varepsilon_{n\mathbf{k}}, \psi_{n\mathbf{k}}, \omega_{\mathbf{q}\nu}, e_{\kappa\alpha,\nu}(\mathbf{q})$  & $\boldsymbol{\epsilon}^\infty, \boldsymbol{Z}$ & $\boldsymbol{Q}$  \\
            &                   & \multicolumn{2}{c|}{$a$ }    & \multicolumn{2}{c|}{$a$ }   &    & & &  \\
LiF         & Fm$\overline{3}$m & \multicolumn{2}{c|}{7.6753}  & \multicolumn{2}{c|}{7.5617} & 45 & 6$\times$6$\times$6
            & 6$\times$6$\times$6 & -- \\
MgO         & Fm$\overline{3}$m & \multicolumn{2}{c|}{8.0368}  & \multicolumn{2}{c|}{7.9635} & 45 & 6$\times$6$\times$6
            & 6$\times$6$\times$6 & --  \\
zb-GaN      & F$\overline{4}$3m & \multicolumn{2}{c|}{8.5981}  & \multicolumn{2}{c|}{8.5032} & 45 & 6$\times$6$\times$6
            & 10$\times$10$\times$10 & 12$\times$12$\times$12 \\
\hline
            &                   & $a$       & $c$              & $a$       & $c$             &    & & &                      \\
wz-GaN      & P6$_3$mc          & 6.0842    & 9.9111           & 6.0173    & 9.8003          & 40 & 8$\times$8$\times$4
            & 12$\times$12$\times$6 & 16$\times$16$\times$8   \\
Li$_2$O$_2$ & P6$_3$/mmc        & 5.9695    & 14.5184           & 5.8852    & 14.3587          & 45 & 8$\times$8$\times$4
            & 8$\times$8$\times$4 & 8$\times$8$\times$4\\
\hline \hline
\end{tabular}
\end{table}

\setcounter{table}{2}

\begin{table}[htbp]
\renewcommand\arraystretch{1.5}
    \centering
    \caption{
    LiF, PBE.
    Dielectric tensor $\boldsymbol{\epsilon}^\infty$ and Born effective charges $\boldsymbol{Z}$.
    }
    \begin{tabular}{llrrrrrrrrr}
\hline \hline
& & \multicolumn{9}{c}{Dielectric tensor components $\epsilon_{ij}$} \\
              &                    &  xx &  xy &  xz &  yx &  yy &  yz &  zx &  zy &  zz \\
 & &{\bf 2.04} & 0.00 & 0.00 & 0.00 & {\bf 2.04} & 0.00 & 0.00 & 0.00 & {\bf 2.04} \\
 \hline
& & \multicolumn{9}{c}{Born effective charges tensor components $Z^{\kappa}_{ij}$~($e$)} \\
Atom $\kappa$ &                    &  xx &  xy &  xz &  yx &  yy &  yz &  zx &  zy &  zz \\
Li & &{\bf 1.05} & 0.00 & 0.00 & 0.00 & {\bf 1.05} & 0.00 & 0.00 & 0.00 & {\bf 1.05} \\
F & &-{\bf 1.05} & 0.00 & 0.00 & 0.00 & -{\bf 1.05} & 0.00 & 0.00 & 0.00 & -{\bf 1.05} \\
\hline \hline
\end{tabular}
\end{table}

\begin{table}[htbp]
\renewcommand\arraystretch{1.5}
    \centering
    \caption{
    LiF, PBEsol.
    Dielectric tensor $\boldsymbol{\epsilon}^\infty$ and Born effective charges $\boldsymbol{Z}$.
    }
    \begin{tabular}{llrrrrrrrrr}
\hline \hline
& & \multicolumn{9}{c}{Dielectric tensor components $\epsilon_{ij}$} \\
              &                    &  xx &  xy &  xz &  yx &  yy &  yz &  zx &  zy &  zz \\
 & &{\bf 2.08} & 0.00 & 0.00 & 0.00 & {\bf 2.08} & 0.00 & 0.00 & 0.00 & {\bf 2.08} \\
 \hline
& & \multicolumn{9}{c}{Born effective charges tensor components $Z^{\kappa}_{ij}$~($e$)} \\
Atom $\kappa$ &                    &  xx &  xy &  xz &  yx &  yy &  yz &  zx &  zy &  zz \\
Li & &{\bf 1.05} & 0.00 & 0.00 & 0.00 & {\bf 1.05} & 0.00 & 0.00 & 0.00 & {\bf 1.05} \\
F & &-{\bf 1.05} & 0.00 & 0.00 & 0.00 & -{\bf 1.05} & 0.00 & 0.00 & 0.00 & -{\bf 1.05} \\
\hline \hline
\end{tabular}
\end{table}

\begin{table}[htbp]
\renewcommand\arraystretch{1.5}
    \centering
    \caption{
    MgO, PBE.
    Dielectric tensor $\boldsymbol{\epsilon}^\infty$ and Born effective charges $\boldsymbol{Z}$.
    }
    \begin{tabular}{llrrrrrrrrr}
\hline \hline
& & \multicolumn{9}{c}{Dielectric tensor components $\epsilon_{ij}$} \\
              &                    &  xx &  xy &  xz &  yx &  yy &  yz &  zx &  zy &  zz \\
 & &{\bf 3.24} & 0.00 & 0.00 & 0.00 & {\bf 3.24} & 0.00 & 0.00 & 0.00 & {\bf 3.24} \\
\hline
& & \multicolumn{9}{c}{Born effective charges tensor components $Z^{\kappa}_{ij}$~($e$)} \\
Atom $\kappa$ &                    &  xx &  xy &  xz &  yx &  yy &  yz &  zx &  zy &  zz \\
Mg & &{\bf 2.00} & 0.00 & 0.00 & 0.00 & {\bf 2.00} & 0.00 & 0.00 & 0.00 & {\bf 2.00} \\
O & &-{\bf 2.00} & 0.00 & 0.00 & 0.00 & -{\bf 2.00} & 0.00 & 0.00 & 0.00 & -{\bf 2.00} \\
\hline \hline
\end{tabular}
\end{table}

\begin{table}[htbp]
\renewcommand\arraystretch{1.5}
    \centering
    \caption{
    MgO, PBEsol.
    Dielectric tensor $\boldsymbol{\epsilon}^\infty$ and Born effective charges $\boldsymbol{Z}$.
    }
    \begin{tabular}{llrrrrrrrrr}
\hline \hline
& & \multicolumn{9}{c}{Dielectric tensor components $\epsilon_{ij}$} \\
              &                    &  xx &  xy &  xz &  yx &  yy &  yz &  zx &  zy &  zz \\
 & &{\bf 3.25} & 0.00 & 0.00 & 0.00 & {\bf 3.25} & 0.00 & 0.00 & 0.00 & {\bf 3.25} \\
 \hline
& & \multicolumn{9}{c}{Born effective charges tensor components $Z^{\kappa}_{ij}$~($e$)} \\
Atom $\kappa$ &                    &  xx &  xy &  xz &  yx &  yy &  yz &  zx &  zy &  zz \\
Mg & &{\bf 1.99} & 0.00 & 0.00 & 0.00 & {\bf 1.99} & 0.00 & 0.00 & 0.00 & {\bf 1.99} \\
O & &-{\bf 1.99} & 0.00 & 0.00 & 0.00 & -{\bf 1.99} & 0.00 & 0.00 & 0.00 & -{\bf 1.99} \\
\hline \hline
\end{tabular}
\end{table}

\begin{table}[htbp]
\renewcommand\arraystretch{1.5}
    \centering
    \caption{
    zb-GaN, PBE.
    Dielectric tensor $\boldsymbol{\epsilon}^\infty$, Born effective charges $\boldsymbol{Z}$ and quadrupoles $\boldsymbol{Q}$.
    }
    \begin{tabular}{llrrrrrrrrr}
\hline \hline
& & \multicolumn{9}{c}{Dielectric tensor components $\epsilon_{ij}$} \\

              &                    &  xx &  xy &  xz &  yx &  yy &  yz &  zx &  zy &  zz \\
              & &{\bf 6.25} & 0.00 & 0.00 & 0.00 & {\bf 6.25} & 0.00 & 0.00 & 0.00 & {\bf 6.25} \\
\hline
& &  \multicolumn{9}{c}{Born effective charges tensor components $Z^{\kappa}_{ij}$~($e$)} \\
Atom $\kappa$ &                    &  xx &  xy &  xz &  yx &  yy &  yz &  zx &  zy &  zz \\
Ga & &{\bf 2.70} & 0.00 & 0.00 & 0.00 & {\bf 2.70} & 0.00 & 0.00 & 0.00 & {\bf 2.70} \\
N & &-{\bf 2.70} & 0.00 & 0.00 & 0.00 & -{\bf 2.70} & 0.00 & 0.00 & 0.00 & -{\bf 2.70} \\
\hline
& & \multicolumn{9}{c}{Quadrupoles tensor components $Q^{\kappa\alpha}_{ij}$~($e\cdot$Bohr)} \\
Atom $\kappa$ & Direction $\alpha$ &  xx &  xy &  xz &  yx &  yy &  yz &  zx &  zy &  zz \\
Ga & x &-{\bf 4.55} & {\bf 3.22} & 0.00 & {\bf 3.22} & 0.00 & 0.00 & 0.00 & 0.00 & {\bf 4.55} \\
 & y &{\bf 3.22} & 0.00 & 0.00 & 0.00 & -{\bf 6.44} & 0.00 & 0.00 & 0.00 & {\bf 3.22} \\
 & z &0.00 & 0.00 & {\bf 4.55} & 0.00 & 0.00 & {\bf 3.22} & {\bf 4.55} & {\bf 3.22} & 0.00 \\
N & x &{\bf 0.24} & -{\bf 0.17} & 0.00 & -{\bf 0.17} & 0.00 & 0.00 & 0.00 & 0.00 & -{\bf 0.24} \\
& y &-{\bf 0.17} & 0.00 & 0.00 & 0.00 & {\bf 0.33} & 0.00 & 0.00 & 0.00 & -{\bf 0.17} \\
& z &0.00 & 0.00 & -{\bf 0.24} & 0.00 & 0.00 & -{\bf 0.17} & -{\bf 0.24} & -{\bf 0.17} & 0.00 \\
\hline \hline
\end{tabular}
\end{table}

\begin{table}[htbp]
\renewcommand\arraystretch{1.5}
    \centering
    \caption{
    zb-GaN, PBEsol.
    Dielectric tensor $\boldsymbol{\epsilon}^\infty$, Born effective charges $\boldsymbol{Z}$ and quadrupoles $\boldsymbol{Q}$.
    }
    \begin{tabular}{llrrrrrrrrr}
\hline \hline
& & \multicolumn{9}{c}{Dielectric tensor components $\epsilon_{ij}$} \\
              &                    &  xx &  xy &  xz &  yx &  yy &  yz &  zx &  zy &  zz \\
 & &{\bf 6.08} & 0.00 & 0.00 & 0.00 & {\bf 6.08} & 0.00 & 0.00 & 0.00 & {\bf 6.08} \\
\hline
& & \multicolumn{9}{c}{Born effective charges tensor components $Z^{\kappa}_{ij}$~($e$)} \\
Atom $\kappa$ &                    &  xx &  xy &  xz &  yx &  yy &  yz &  zx &  zy &  zz \\
Ga & &{\bf 2.67} & 0.00 & 0.00 & 0.00 & {\bf 2.67} & 0.00 & 0.00 & 0.00 & {\bf 2.67} \\
N & &-{\bf 2.67} & 0.00 & 0.00 & 0.00 & -{\bf 2.67} & 0.00 & 0.00 & 0.00 & -{\bf 2.67} \\
\hline
& & \multicolumn{9}{c}{Quadrupoles tensor components $Q^{\kappa\alpha}_{ij}$~($e\cdot$Bohr)} \\
Atom $\kappa$ & Direction $\alpha$ &  xx &  xy &  xz &  yx &  yy &  yz &  zx &  zy &  zz \\
Ga & x &-{\bf 4.27} & {\bf 3.02} & 0.00 & {\bf 3.02} & 0.00 & 0.00 & 0.00 & 0.00 & {\bf 4.27} \\
& y &{\bf 3.02} & 0.00 & 0.00 & 0.00 & -{\bf 6.03} & 0.00 & 0.00 & 0.00 & {\bf 3.02} \\
 & z &0.00 & 0.00 & {\bf 4.27} & 0.00 & 0.00 & {\bf 3.02} & {\bf 4.27} & {\bf 3.02} & 0.00 \\
N & x &{\bf 0.17} & -{\bf 0.12} & 0.00 & -{\bf 0.12} & 0.00 & 0.00 & 0.00 & 0.00 & -{\bf 0.17} \\
& y &-{\bf 0.12} & 0.00 & 0.00 & 0.00 & {\bf 0.24} & 0.00 & 0.00 & 0.00 & -{\bf 0.12} \\
& z &0.00 & 0.00 & -{\bf 0.17} & 0.00 & 0.00 & -{\bf 0.12} & -{\bf 0.17} & -{\bf 0.12} & 0.00 \\
\hline \hline
\end{tabular}
\end{table}

\begin{table}[htbp]
\renewcommand\arraystretch{1.5}
    \centering
    \caption{
    wz-GaN, PBE.
    Dielectric tensor $\boldsymbol{\epsilon}^\infty$, Born effective charges $\boldsymbol{Z}$ and quadrupoles $\boldsymbol{Q}$.
    }
    \begin{tabular}{llrrrrrrrrr}
\hline \hline
& & \multicolumn{9}{c}{Dielectric tensor components $\epsilon_{ij}$} \\
              &                    &  xx &  xy &  xz &  yx &  yy &  yz &  zx &  zy &  zz \\
 & &{\bf 5.97} & 0.00 & 0.00 & 0.00 & {\bf 5.97} & 0.00 & 0.00 & 0.00 & {\bf 6.18} \\
\hline
& & \multicolumn{9}{c}{Born effective charges tensor components $Z^{\kappa}_{ij}$~($e$)} \\
Atom $\kappa$ &                    &  xx &  xy &  xz &  yx &  yy &  yz &  zx &  zy &  zz \\
Ga$_1$ & &{\bf 2.65} & 0.00 & 0.00 & 0.00 & {\bf 2.65} & 0.00 & 0.00 & 0.00 & {\bf 2.79} \\
Ga$_2$ & &{\bf 2.65} & 0.00 & 0.00 & 0.00 & {\bf 2.65} & 0.00 & 0.00 & 0.00 & {\bf 2.79} \\
N$_1$ & &-{\bf 2.65} & 0.00 & 0.00 & 0.00 & -{\bf 2.65} & 0.00 & 0.00 & 0.00 & -{\bf 2.79} \\
N$_1$ & &-{\bf 2.65} & 0.00 & 0.00 & 0.00 & -{\bf 2.65} & 0.00 & 0.00 & 0.00 & -{\bf 2.79} \\
\hline
& & \multicolumn{9}{c}{Quadrupoles tensor components $Q^{\kappa\alpha}_{ij}$~($e\cdot$Bohr)} \\
Atom $\kappa$ & Direction $\alpha$ &  xx &  xy &  xz &  yx &  yy &  yz &  zx &  zy &  zz \\
Ga$_1$ & x &0.00 & -{\bf 4.29} & -{\bf 2.70} & -{\bf 4.29} & 0.00 & 0.00 & -{\bf 2.70} & 0.00 & 0.00 \\
& y &-{\bf 4.29} & 0.00 & 0.00 & 0.00 & {\bf 4.29} & -{\bf 2.70} & 0.00 & -{\bf 2.70} & 0.00 \\
& z &-{\bf 2.77} & 0.00 & 0.00 & 0.00 & -{\bf 2.77} & 0.00 & 0.00 & 0.00 & {\bf 5.76} \\
Ga$_2$ & x &0.00 & {\bf 4.29} & -{\bf 2.70} & {\bf 4.29} & 0.00 & 0.00 & -{\bf 2.70} & 0.00 & 0.00 \\
 & y &{\bf 4.29} & 0.00 & 0.00 & 0.00 & -{\bf 4.29} & -{\bf 2.70} & 0.00 & -{\bf 2.70} & 0.00 \\
 & z &-{\bf 2.77} & 0.00 & 0.00 & 0.00 & -{\bf 2.77} & 0.00 & 0.00 & 0.00 & {\bf 5.76} \\
N$_1$ & x &0.00 & {\bf 1.21} & -{\bf 0.02} & {\bf 1.21} & 0.00 & 0.00 & -{\bf 0.02} & 0.00 & 0.00 \\
 & y &{\bf 1.21} & 0.00 & 0.00 & 0.00 & -{\bf 1.21} & -{\bf 0.02} & 0.00 & -{\bf 0.02} & 0.00 \\
 & z &-{\bf 0.07} & 0.00 & 0.00 & 0.00 & -{\bf 0.07} & 0.00 & 0.00 & 0.00 & -{\bf 0.50} \\
N$_2$ & x &0.00 & -{\bf 1.21} & -{\bf 0.02} & -{\bf 1.21} & 0.00 & 0.00 & -{\bf 0.02} & 0.00 & 0.00 \\
 & y &-{\bf 1.21} & 0.00 & 0.00 & 0.00 & {\bf 1.21} & -{\bf 0.02} & 0.00 & -{\bf 0.02} & 0.00 \\
 & z &-{\bf 0.07} & 0.00 & 0.00 & 0.00 & -{\bf 0.07} & 0.00 & 0.00 & 0.00 & -{\bf 0.50} \\
\hline \hline
\end{tabular}
\end{table}

\begin{table}[htbp]
\renewcommand\arraystretch{1.5}
    \centering
    \caption{
    wz-GaN, PBEsol.
    Dielectric tensor $\boldsymbol{\epsilon}^\infty$, Born effective charges $\boldsymbol{Z}$ and quadrupoles $\boldsymbol{Q}$.
    }
    \begin{tabular}{llrrrrrrrrr}
\hline \hline
& & \multicolumn{9}{c}{Dielectric tensor components $\epsilon_{ij}$} \\
              &                    &  xx &  xy &  xz &  yx &  yy &  yz &  zx &  zy &  zz \\
 & &{\bf 5.85} & 0.00 & 0.00 & 0.00 & {\bf 5.85} & 0.00 & 0.00 & 0.00 & {\bf 6.05} \\
\hline
& & \multicolumn{9}{c}{Born effective charges tensor components $Z^{\kappa}_{ij}$~($e$)} \\
Atom $\kappa$ &                    &  xx &  xy &  xz &  yx &  yy &  yz &  zx &  zy &  zz \\
Ga$_1$ & &{\bf 2.63} & 0.00 & 0.00 & 0.00 & {\bf 2.63} & 0.00 & 0.00 & 0.00 & {\bf 2.77} \\
Ga$_2$ & &{\bf 2.63} & 0.00 & 0.00 & 0.00 & {\bf 2.63} & 0.00 & 0.00 & 0.00 & {\bf 2.77} \\
N$_1$ & &-{\bf 2.63} & 0.00 & 0.00 & 0.00 & -{\bf 2.63} & 0.00 & 0.00 & 0.00 & -{\bf 2.77} \\
N$_1$ & &-{\bf 2.63} & 0.00 & 0.00 & 0.00 & -{\bf 2.63} & 0.00 & 0.00 & 0.00 & -{\bf 2.77} \\
\hline
& & \multicolumn{9}{c}{Quadrupoles tensor components $Q^{\kappa\alpha}_{ij}$~($e\cdot$Bohr)} \\
Atom $\kappa$ & Direction $\alpha$ &  xx &  xy &  xz &  yx &  yy &  yz &  zx &  zy &  zz \\
Ga$_1$ & x &0.00 & -{\bf 4.06} & -{\bf 2.53} & -{\bf 4.06} & 0.00 & 0.00 & -{\bf 2.53} & 0.00 & 0.00 \\
& y &-{\bf 4.06} & 0.00 & 0.00 & 0.00 & {\bf 4.06} & -{\bf 2.53} & 0.00 & -{\bf 2.53} & 0.00 \\
& z &-{\bf 2.60} & 0.00 & 0.00 & 0.00 & -{\bf 2.60} & 0.00 & 0.00 & 0.00 & {\bf 5.38} \\
Ga$_2$ & x &0.00 & {\bf 4.06} & -{\bf 2.53} & {\bf 4.06} & 0.00 & 0.00 & -{\bf 2.53} & 0.00 & 0.00 \\
& y &{\bf 4.06} & 0.00 & 0.00 & 0.00 & -{\bf 4.06} & -{\bf 2.53} & 0.00 & -{\bf 2.53} & 0.00 \\
& z &-{\bf 2.60} & 0.00 & 0.00 & 0.00 & -{\bf 2.60} & 0.00 & 0.00 & 0.00 & {\bf 5.38} \\
N$_1$ & x &0.00 & {\bf 1.20} & -{\bf 0.05} & {\bf 1.20} & 0.00 & 0.00 & -{\bf 0.05} & 0.00 & 0.00 \\
& y &{\bf 1.20} & 0.00 & 0.00 & 0.00 & -{\bf 1.20} & -{\bf 0.05} & 0.00 & -{\bf 0.05} & 0.00 \\
& z &-{\bf 0.11} & 0.00 & 0.00 & 0.00 & -{\bf 0.11} & 0.00 & 0.00 & 0.00 & -{\bf 0.43} \\
N$_2$ & x &0.00 & -{\bf 1.20} & -{\bf 0.05} & -{\bf 1.20} & 0.00 & 0.00 & -{\bf 0.05} & 0.00 & 0.00 \\
& y &-{\bf 1.20} & 0.00 & 0.00 & 0.00 & {\bf 1.20} & -{\bf 0.05} & 0.00 & -{\bf 0.05} & 0.00 \\
& z &-{\bf 0.11} & 0.00 & 0.00 & 0.00 & -{\bf 0.11} & 0.00 & 0.00 & 0.00 & -{\bf 0.43} \\
\hline \hline
\end{tabular}
\end{table}

\begin{table}[htbp]
\renewcommand\arraystretch{1.5}
    \centering
    \caption{
    Li$_2$O$_2$, PBE.
    Dielectric tensor $\boldsymbol{\epsilon}^\infty$, Born effective charges $\boldsymbol{Z}$ and quadrupoles $\boldsymbol{Q}$.
    }
    \begin{tabular}{llrrrrrrrrr}
\hline \hline
& & \multicolumn{9}{c}{Dielectric tensor components $\epsilon_{ij}$} \\
              &                    &  xx &  xy &  xz &  yx &  yy &  yz &  zx &  zy &  zz \\
 & &{\bf 2.73} & 0.00 & 0.00 & 0.00 & {\bf 2.73} & 0.00 & 0.00 & 0.00 & {\bf 3.92} \\
\hline
& & \multicolumn{9}{c}{Born effective charges tensor components $Z^{\kappa}_{ij}$~($e$)} \\
Atom $\kappa$ &                    &  xx &  xy &  xz &  yx &  yy &  yz &  zx &  zy &  zz \\
Li$_1$ & &{\bf 1.08} & 0.00 & 0.00 & 0.00 & {\bf 1.08} & 0.00 & 0.00 & 0.00 & {\bf 1.27} \\
Li$_2$ & &{\bf 1.08} & 0.00 & 0.00 & 0.00 & {\bf 1.08} & 0.00 & 0.00 & 0.00 & {\bf 1.27} \\
Li$_3$ & &{\bf 1.06} & 0.00 & 0.00 & 0.00 & {\bf 1.06} & 0.00 & 0.00 & 0.00 & {\bf 0.69} \\
Li$_4$ & &{\bf 1.06} & 0.00 & 0.00 & 0.00 & {\bf 1.06} & 0.00 & 0.00 & 0.00 & {\bf 0.69} \\
O$_1$ & &-{\bf 1.07} & 0.00 & 0.00 & 0.00 & -{\bf 1.07} & 0.00 & 0.00 & 0.00 & -{\bf 0.98} \\
O$_2$ & &-{\bf 1.07} & 0.00 & 0.00 & 0.00 & -{\bf 1.07} & 0.00 & 0.00 & 0.00 & -{\bf 0.98} \\
O$_3$ & &-{\bf 1.07} & 0.00 & 0.00 & 0.00 & -{\bf 1.07} & 0.00 & 0.00 & 0.00 & -{\bf 0.98} \\
O$_4$ & &-{\bf 1.07} & 0.00 & 0.00 & 0.00 & -{\bf 1.07} & 0.00 & 0.00 & 0.00 & -{\bf 0.98} \\
\hline
& & \multicolumn{9}{c}{Quadrupoles tensor components $Q^{\kappa\alpha}_{ij}$~($e\cdot$Bohr)} \\
Atom $\kappa$ & Direction $\alpha$ &  xx &  xy &  xz &  yx &  yy &  yz &  zx &  zy &  zz \\
Li$_1$ & x &0.00 & 0.00 & 0.00 & 0.00 & 0.00 & 0.00 & 0.00 & 0.00 & 0.00 \\
 & y &0.00 & 0.00 & 0.00 & 0.00 & 0.00 & 0.00 & 0.00 & 0.00 & 0.00 \\
 & z &0.00 & 0.00 & 0.00 & 0.00 & 0.00 & 0.00 & 0.00 & 0.00 & 0.00 \\
Li$_2$ & x &0.00 & 0.00 & 0.00 & 0.00 & 0.00 & 0.00 & 0.00 & 0.00 & 0.00 \\
 & y &0.00 & 0.00 & 0.00 & 0.00 & 0.00 & 0.00 & 0.00 & 0.00 & 0.00 \\
 & z &0.00 & 0.00 & 0.00 & 0.00 & 0.00 & 0.00 & 0.00 & 0.00 & 0.00 \\
Li$_3$ & x &0.00 & {\bf 1.10} & 0.00 & {\bf 1.10} & 0.00 & 0.00 & 0.00 & 0.00 & 0.00 \\
 & y &{\bf 1.10} & 0.00 & 0.00 & 0.00 & -{\bf 1.10} & 0.00 & 0.00 & 0.00 & 0.00 \\
 & z &0.00 & 0.00 & 0.00 & 0.00 & 0.00 & 0.00 & 0.00 & 0.00 & 0.00 \\
Li$_4$ & x &0.00 & -{\bf 1.10} & 0.00 & -{\bf 1.10} & 0.00 & 0.00 & 0.00 & 0.00 & 0.00 \\
 & y &-{\bf 1.10} & 0.00 & 0.00 & 0.00 & {\bf 1.10} & 0.00 & 0.00 & 0.00 & 0.00 \\
 & z &0.00 & 0.00 & 0.00 & 0.00 & 0.00 & 0.00 & 0.00 & 0.00 & 0.00 \\
O$_1$ & x &0.00 & -{\bf 0.05} & -{\bf 1.04} & -{\bf 0.05} & 0.00 & 0.00 & -{\bf 1.04} & 0.00 & 0.00 \\
 & y &-{\bf 0.05} & 0.00 & 0.00 & 0.00 & {\bf 0.05} & -{\bf 1.04} & 0.00 & -{\bf 1.04} & 0.00 \\
 & z &-{\bf 0.46} & 0.00 & 0.00 & 0.00 & -{\bf 0.46} & 0.00 & 0.00 & 0.00 & -{\bf 4.91} \\
O$_2$ & x &0.00 & {\bf 0.05} & -{\bf 1.04} & {\bf 0.05} & 0.00 & 0.00 & -{\bf 1.04} & 0.00 & 0.00 \\
 & y &{\bf 0.05} & 0.00 & 0.00 & 0.00 & -{\bf 0.05} & -{\bf 1.04} & 0.00 & -{\bf 1.04} & 0.00 \\
 & z &-{\bf 0.46} & 0.00 & 0.00 & 0.00 & -{\bf 0.46} & 0.00 & 0.00 & 0.00 & -{\bf 4.91} \\
O$_3$ & x &0.00 & {\bf 0.05} & {\bf 1.04} & {\bf 0.05} & 0.00 & 0.00 & {\bf 1.04} & 0.00 & 0.00 \\
 & y &{\bf 0.05} & 0.00 & 0.00 & 0.00 & -{\bf 0.05} & {\bf 1.04} & 0.00 & {\bf 1.04} & 0.00 \\
 & z &{\bf 0.46} & 0.00 & 0.00 & 0.00 & {\bf 0.46} & 0.00 & 0.00 & 0.00 & {\bf 4.91} \\
O$_4$ & x &0.00 & -{\bf 0.05} & {\bf 1.04} & -{\bf 0.05} & 0.00 & 0.00 & {\bf 1.04} & 0.00 & 0.00 \\
 & y &-{\bf 0.05} & 0.00 & 0.00 & 0.00 & {\bf 0.05} & {\bf 1.04} & 0.00 & {\bf 1.04} & 0.00 \\
 & z &{\bf 0.46} & 0.00 & 0.00 & 0.00 & {\bf 0.46} & 0.00 & 0.00 & 0.00 & {\bf 4.91} \\
\hline \hline
\end{tabular}
\end{table}

\begin{table}[htbp]
\renewcommand\arraystretch{1.5}
    \centering
    \caption{
    Li$_2$O$_2$, PBEsol.
    Dielectric tensor $\boldsymbol{\epsilon}^\infty$, Born effective charges $\boldsymbol{Z}$ and quadrupoles $\boldsymbol{Q}$.
    }
    \begin{tabular}{llrrrrrrrrr}
\hline \hline
& & \multicolumn{9}{c}{Dielectric tensor components $\epsilon_{ij}$} \\
              &                    &  xx &  xy &  xz &  yx &  yy &  yz &  zx &  zy &  zz \\
 & &{\bf 2.78} & 0.00 & 0.00 & 0.00 & {\bf 2.78} & 0.00 & 0.00 & 0.00 & {\bf 3.96} \\
\hline
& & \multicolumn{9}{c}{Born effective charges tensor components $Z^{\kappa}_{ij}$~($e$)} \\
Atom $\kappa$ &                    &  xx &  xy &  xz &  yx &  yy &  yz &  zx &  zy &  zz \\
Li$_1$ & &{\bf 1.07} & 0.00 & 0.00 & 0.00 & {\bf 1.07} & 0.00 & 0.00 & 0.00 & {\bf 1.25} \\
Li$_2$ & &{\bf 1.07} & 0.00 & 0.00 & 0.00 & {\bf 1.07} & 0.00 & 0.00 & 0.00 & {\bf 1.25} \\
Li$_3$ & &{\bf 1.05} & 0.00 & 0.00 & 0.00 & {\bf 1.05} & 0.00 & 0.00 & 0.00 & {\bf 0.69} \\
Li$_4$ & &{\bf 1.05} & 0.00 & 0.00 & 0.00 & {\bf 1.05} & 0.00 & 0.00 & 0.00 & {\bf 0.69} \\
O$_1$ & &-{\bf 1.06} & 0.00 & 0.00 & 0.00 & -{\bf 1.06} & 0.00 & 0.00 & 0.00 & -{\bf 0.97} \\
O$_2$ & &-{\bf 1.06} & 0.00 & 0.00 & 0.00 & -{\bf 1.06} & 0.00 & 0.00 & 0.00 & -{\bf 0.97} \\
O$_3$ & &-{\bf 1.06} & 0.00 & 0.00 & 0.00 & -{\bf 1.06} & 0.00 & 0.00 & 0.00 & -{\bf 0.97} \\
O$_4$ & &-{\bf 1.06} & 0.00 & 0.00 & 0.00 & -{\bf 1.06} & 0.00 & 0.00 & 0.00 & -{\bf 0.97} \\
\hline
& & \multicolumn{9}{c}{Quadrupoles tensor components $Q^{\kappa\alpha}_{ij}$~($e\cdot$Bohr)} \\
Atom $\kappa$ & Direction $\alpha$ &  xx &  xy &  xz &  yx &  yy &  yz &  zx &  zy &  zz \\
Li$_1$ & x &0.00 & 0.00 & 0.00 & 0.00 & 0.00 & 0.00 & 0.00 & 0.00 & 0.00 \\
& y &0.00 & 0.00 & 0.00 & 0.00 & 0.00 & 0.00 & 0.00 & 0.00 & 0.00 \\
& z &0.00 & 0.00 & 0.00 & 0.00 & 0.00 & 0.00 & 0.00 & 0.00 & 0.00 \\
Li$_2$  & x &0.00 & 0.00 & 0.00 & 0.00 & 0.00 & 0.00 & 0.00 & 0.00 & 0.00 \\
& y &0.00 & 0.00 & 0.00 & 0.00 & 0.00 & 0.00 & 0.00 & 0.00 & 0.00 \\
& z &0.00 & 0.00 & 0.00 & 0.00 & 0.00 & 0.00 & 0.00 & 0.00 & 0.00 \\
Li$_3$  & x &0.00 & -{\bf 1.06} & 0.00 & -{\bf 1.06} & 0.00 & 0.00 & 0.00 & 0.00 & 0.00 \\
& y &-{\bf 1.06} & 0.00 & 0.00 & 0.00 & {\bf 1.06} & 0.00 & 0.00 & 0.00 & 0.00 \\
& z &0.00 & 0.00 & 0.00 & 0.00 & 0.00 & 0.00 & 0.00 & 0.00 & 0.00 \\
Li$_4$  & x &0.00 & {\bf 1.06} & 0.00 & {\bf 1.06} & 0.00 & 0.00 & 0.00 & 0.00 & 0.00 \\
& y &{\bf 1.06} & 0.00 & 0.00 & 0.00 & -{\bf 1.06} & 0.00 & 0.00 & 0.00 & 0.00 \\
& z &0.00 & 0.00 & 0.00 & 0.00 & 0.00 & 0.00 & 0.00 & 0.00 & 0.00 \\
O$_1$  & x &0.00 & {\bf 0.03} & -{\bf 1.05} & {\bf 0.03} & 0.00 & 0.00 & -{\bf 1.05} & 0.00 & 0.00 \\
& y &{\bf 0.03} & 0.00 & 0.00 & 0.00 & -{\bf 0.03} & -{\bf 1.05} & 0.00 & -{\bf 1.05} & 0.00 \\
& z &-{\bf 0.45} & 0.00 & 0.00 & 0.00 & -{\bf 0.45} & 0.00 & 0.00 & 0.00 & -{\bf 5.11} \\
O$_2$  & x &0.00 & -{\bf 0.03} & -{\bf 1.05} & -{\bf 0.03} & 0.00 & 0.00 & -{\bf 1.05} & 0.00 & 0.00 \\
& y &-{\bf 0.03} & 0.00 & 0.00 & 0.00 & {\bf 0.03} & -{\bf 1.05} & 0.00 & -{\bf 1.05} & 0.00 \\
& z &-{\bf 0.45} & 0.00 & 0.00 & 0.00 & -{\bf 0.45} & 0.00 & 0.00 & 0.00 & -{\bf 5.11} \\
O$_3$  & x &0.00 & -{\bf 0.03} & {\bf 1.05} & -{\bf 0.03} & 0.00 & 0.00 & {\bf 1.05} & 0.00 & 0.00 \\
& y &-{\bf 0.03} & 0.00 & 0.00 & 0.00 & {\bf 0.03} & {\bf 1.05} & 0.00 & {\bf 1.05} & 0.00 \\
& z &{\bf 0.45} & 0.00 & 0.00 & 0.00 & {\bf 0.45} & 0.00 & 0.00 & 0.00 & {\bf 5.11} \\
O$_4$  & x &0.00 & {\bf 0.03} & {\bf 1.05} & {\bf 0.03} & 0.00 & 0.00 & {\bf 1.05} & 0.00 & 0.00 \\
& y &{\bf 0.03} & 0.00 & 0.00 & 0.00 & -{\bf 0.03} & {\bf 1.05} & 0.00 & {\bf 1.05} & 0.00 \\
& z &{\bf 0.45} & 0.00 & 0.00 & 0.00 & {\bf 0.45} & 0.00 & 0.00 & 0.00 & {\bf 5.11} \\
\hline \hline
\end{tabular}
\end{table}

\begin{table}[htbp]
\renewcommand\arraystretch{1.5}
    \centering
    \caption{
    Average theoretical speedup $s_{\Delta \varepsilon}$ for large polaron calculation with energy filtering $\Delta \varepsilon$ employed in this work.
    $s_{\Delta \varepsilon}$ were calculated using Eq.~(42) of the main text for various $\mathbf{k}$-meshes, and averaged to give the estimated speedup with standard deviation.
    }
    \begin{tabular}{lccc}
\hline \hline
Materials & $\mathbf{k}$-mesh & $\Delta \varepsilon$~(eV) & $s_{\Delta \varepsilon}$~(-) \\
\hline
LiF       & 10$\times$10$\times$10 $\to$ 24$\times$24$\times$24 &  2.50 &               3.3~$\pm$~0.1  \\
MgO       & 80$\times$80$\times$80 $\to$ 95$\times$95$\times$95 &                     0.25 &             900.6~$\pm$~29.4 \\
wz-GaN    & 70$\times$70$\times$35 $\to$ 90$\times$90$\times$45 &                    0.09 &             136.9~$\pm$~2.6  \\
zb-GaN    & 75$\times$75$\times$75 $\to$ 85$\times$85$\times$85 &                    0.09 &             199.8~$\pm$~3.3  \\
\hline \hline
\end{tabular}
\end{table}

\begin{table}[htbp]
\renewcommand\arraystretch{1.5}
    \centering
    \caption{
    MgO, PBE and PBEsol.
    Macroscopic parameters required for the Fr\"ohlich model:
    electron effective mass $m^*_{e}$,
    static and high-frequency dielectric constants $\epsilon^\infty$ and $\epsilon^0$,
    LO phonon frequency $\omega_{\rm LO}$ at at $\mathbf{q} \to \Gamma$ ,
    and corresponding Fr\"ohlich coupling constant $\alpha$.
    }
    \begin{tabular}{lccccc}
\hline \hline
XC     & $m^*_e$ & $\epsilon^\infty$ & 
$\epsilon^0$ & $\omega_{\rm LO}$ (meV) & $\alpha$ \\
\hline
PBE    &   0.339 &               3.244  &          11.174 &                    84.3 &    1.617 \\
PBEsol &   0.344 &               3.250  &          10.462 &                    86.0 &    1.565 \\
\hline \hline
\end{tabular}
\end{table}
